\documentstyle[12pt]{article}

\newcommand{\tr}{\mbox{Tr} }
\newcommand{\ket}[1]{\left | #1 \right \rangle}
\newcommand{\bra}[1]{\left \langle #1 \right |}
\newcommand{\amp}[2]{\left \langle #1 \left | #2 \right. \right \rangle}

\newcommand{\proj}[1]{\ket{#1} \! \bra{#1}}

\newcommand{\superop}{\mbox{\$}}
\newcommand{\unity}{\mbox{\bf I}}
\newcommand{\hilbert}[1]{{\cal H}_{#1}}

\begin{document}

\begin{center}
{\Large \bf Sending entanglement through\\
noisy quantum channels} \bigskip \\
{\large Benjamin Schumacher} \medskip \\
Theoretical Astrophysics, T-6  MS B288\\
Los Alamos National Laboratory, Los Alamos, NM 87545 \medskip \\
Permanent address:  Department of Physics, Kenyon College,\\
Gambier, Ohio  43022
\end{center}

\bigskip
\bigskip

\section*{Abstract}

This paper addresses some general questions of quantum information theory
arising from the transmission of quantum entanglement through (possibly
noisy) quantum channels.  A pure entangled state is prepared of a pair
of systems $R$ and $Q$, after which $Q$ is subjected to a dynamical
evolution given by the superoperator $\superop^{Q}$.  Two interesting
quantities can be defined for this process:  the entanglement fidelity 
$F_{e}$ and the entropy production $S_{e}$.  It turns out that neither 
of these quantities depends in any way on the system $R$, but only on 
the initial state and dynamical evolution of $Q$.  $F_{e}$ and $S_{e}$ are
related to various other fidelities and entropies, and are connected by
an inequality reminiscent of the Fano inequality of classical information
theory.  Some insight can be gained from these techniques into the
security of quantum cryptographic protocols and the nature of quantum
error-correcting codes.

\vfill

PACS numbers:  03.65Bz, 05.30.-d, 89.70.+c

\pagebreak

\section{Introduction}

In recent years, considerable progress has been made toward developing
a general quantum theory of information
\cite{chb-phystoday}, analogous to classical information
theory founded by Shannon \cite{shannon}.  Distinctively quantum-mechanical
notions of coding \cite{qcoding} and channel fidelity \cite{qfidelity}
have been developed, and the role of entangled states in storing and
transferring quantum information has been explored \cite{entang}.
Recently, the study of noisy quantum channels has yielded 
important new results about quantum error-correcting codes \cite{qerror} 
and the purification of noisy entangled states \cite{purify}.

The aim of this paper is to further clarify our understanding of noisy
quantum channels by defining and exploiting new notions of fidelity and
entropy associated with the quantum transmission process.  
These new quantities are based on an analysis of the
transmission of entangled states through the noisy channel,  although
(as we shall see) the use of entanglement is not essential to their
definition.  A number of applications of these ideas will be outlined.

Here is the general situation that we will consider.  Suppose $R$ and $Q$
are two quantum systems, and $Q$ is described by a Hilbert space $\hilbert{Q}$
of finite dimension $d$.  Initially the joint system $RQ$ is prepared in a 
pure entangled state $\ket{\Psi^{RQ}}$.  
The system $R$ is dynamically isolated
and has a zero internal Hamiltonian, while the system $Q$ undergoes some 
evolution that possibly involves interaction with the environment $E$.
The evolution of $Q$ might, for example, represent a coding, transmission, and
decoding process via some quantum channel for the quantum information in $Q$.
The final state of $RQ$ is possibly mixed, and is described by the 
density operator $\rho^{RQ'}$.

The {\em fidelity} of this process is 
$F_{e} =  \bra{\Psi^{RQ}} \rho^{RQ'} \ket{\Psi^{RQ}}$, 
which is the probability that the final state $\rho^{RQ'}$ would pass
a test checking whether it agreed with the initial state $\ket{\Psi^{RQ}}$.
(This imagined test would be a measurement of a joint observable on $RQ$.)
$F_{e}$ measures how successfully the quantum channel preserves the
entanglement of $Q$ with the ``reference system'' $R$.

We will demonstrate three important results.
First, the fidelity $F_{e}$ can be defined entirely in terms of the
initial state and evolution of the system $Q$.  Furthermore,
$F_{e} \leq \bar{F}$, where $\bar{F}$ is the average fidelity 
when the channel carries one of an ensemble of pure states of $Q$ described by
$\rho^{Q} = \tr_{R} \proj{\Psi^{RQ}}$.  Thus, channels which
can convey entanglement faithfully will also convey ensembles
of pure states faithfully.

Second, there exists a quantity $S_{e}$ called {\em entropy production}, also
defined in terms of the internal properties of the system $Q$.
This quantity can be viewed as the amount of information that
is exchanged with the environment during the interaction of $Q$ and $E$,
and it characterizes the amount of ``quantum noise'' in the evolution of $Q$.

Finally, we will find an inequality (resembling the Fano inequality of
classical information theory) that bounds $F_{e}$ in terms of
the dimension $d$ and the entropy production $S_{e}$ in $Q$.
In other words, the faithfulness of $Q$'s dynamical evolution in preserving
entanglement is limited by the amount of information 
that is exchanged with the environment.

The Appendix uses some ideas from the paper to give a derivation of 
two representation theorems for trace-preserving, completely positive 
maps, which are the most general descriptions for quantum dynamical 
evolutions \cite{cpmaps}.

Throughout this paper, the systems relevant to a particular vector,
operator, or superoperator will be indicated by a superscript.  Thus, 
$\ket{\psi^{Q}}$ is a state vector for the system $Q$, while $A^{RQ}$ is
an operator acting on $\hilbert{RQ} = \hilbert{R} \otimes \hilbert{Q}$.
(If no superscript is given, the quantum system is supposed to be generic.)  
A prime symbol ( $'$ ) denotes that a particular state or density 
operator arises as a result of some dynamical evolution.
A tilde symbol ( $\sim$ ) is usually present when a particular state
vector or operator is not normalized, so that $\amp{\psi}{\psi} = 1$,
but $\amp{\tilde{\psi}}{\tilde{\psi}} \neq 1$ in general.

\section{Channel Dynamics}

\subsection{Completely positive maps}

Imagine that the system $Q$ is prepared in an initial state $\rho^{Q}$ and
then subjected to some dynamical process, after which the state is $\rho^{Q'}$.
The dynamical process is described by a map $\superop^{Q}$, so that the 
evolution is
\begin{displaymath}
	\rho^{Q} \longrightarrow \rho^{Q'} = \superop^{Q}(\rho^{Q}) .
\end{displaymath}
In the most general case, the map $\superop^{Q}$ must be a trace-preserving,
completely positive linear map \cite{cpmaps}.  In other words,
\begin{itemize}
	\item  $\superop^{Q}$ must be linear in the density operators.  
		That is, if $\rho^{Q} = p_{1} \rho^{Q}_{1} + 
		p_{2} \rho^{Q}_{2}$, then
		\begin{eqnarray*}
	\superop^{Q}(\rho^{Q'}) 
		& = & p_{1} \rho^{Q'}_{1} + p_{2} \rho^{Q'}_{2} \\
			 & = & p_{1} \left(\superop^{Q}(\rho^{Q}_{1}) \right)
			   +  p_{2} \left(\superop^{Q}(\rho^{Q}_{2}) \right).
		\end{eqnarray*}
		A probabilistic mixture of inputs to 
		$\superop^{Q}$ leads to a probabilistic mixture of outputs.
		This means that $\superop^{Q}$ must be a {\em superoperator},
		that is, a linear operator acting on the space of linear
		operators (e.g., density operators) on $\hilbert{Q}$.
	\item  $\superop^{Q}$ must be trace-preserving, so that
		$\tr \, \rho^{Q'} = \tr \, \rho^{Q} = 1$.
	\item  $\superop^{Q}$ must be positive.  This means that if $\rho^{Q}$
		is positive\footnote{We will use the term ``positive'' 
		to refer generically to operators that are {\em positive 
		semi-definite}---i.e., those that are Hermitian and 
		have no negative eigenvalues.} then 
		$\rho^{Q'} = \superop^{Q}(\rho^{Q})$ must be positive.
\end{itemize}
These three conditions mean that the superoperator $\superop^{Q}$ 
takes normalized density operators to normalized density operators in a
reasonable way.  The requirement of {\em complete} positivity is somewhat
more subtle.
\begin{itemize}
	\item  $\superop^{Q}$ must be completely positive.  That is, suppose
		we extend the evolution superoperator $\superop^{Q}$ in a
		trivial way to an evolution superoperator for a compound
		system $RQ$, yielding  $\unity^{R} \otimes \superop^{Q}$,
		where $\unity^{R}$ is the identity 
		superoperator on $R$ states.
		Physically, this means adjoining a system $R$ which has
		trivial dynamics (no state of $R$ is changed) and which
		does not interact with $Q$.  $\superop^{Q}$ is completely
		positive if, for all such trivial extensions, the resulting
		superoperator $\unity^{R} \otimes \superop^{Q}$ is positive.
\end{itemize}
A completely positive map is not only a reasonable map from density
operators to density operators for $Q$, but it is {\em extensible} in a
trivial way to a reasonable map from density operators to density operators 
on any larger system $RQ$.  Since we cannot exclude {\em a priori} that
our system $Q$ is in fact initially entangled with some distant isolated 
system $R$, any acceptable $\superop^{Q}$ had better satisfy this
condition.

\subsection{Representations of $\superop^{Q}$}

Completely positive, trace-preserving linear maps obviously include all
unitary evolutions of the state:  
$\rho^{Q'} = U^{Q} \rho^{Q} {U^{Q}}^{\dagger}$.
They also include unitary evolutions involving interactions with an
external system.  Suppose we consider an environment system $E$ that is
initially in the pure state $\ket{0^{E}}$.  Then we could have
\begin{equation}
	\superop^{Q}(\rho^{Q}) = \tr_{E} \, U^{QE} \left ( 
				 \rho^{Q} \otimes \proj{0^{E}}
				 \right )  {U^{QE}}^{\dagger}     
				\label{bigunitary}
\end{equation}
where $U^{QE}$ is some arbitrary unitary evolution on the joint system $QE$.
This map is also trace-preserving and completely positive.

If we can write a superoperator $\superop^{Q}$ as a unitary evolution on an
extended system $QE$ followed by a partial trace over $E$, we say that
we have a ``unitary representation'' of the superoperator.  Such a
representation is not unique, since many different unitary
operators $U^{QE}$ will lead to the same $\superop^{Q}$.

Another useful sort of representation for completely positive maps 
employs only operators on $\hilbert{Q}$.
Let $A^{Q}_{\mu}$ be a collection of such operators 
indexed by $\mu$.  Then the map
$\superop^{Q}$ given by
\begin{equation}
	\superop^{Q}(\rho^{Q}) = \sum_{\mu} A^{Q}_{\mu} \rho^{Q} 
		{A^{Q}_{\mu}}^{\dagger}
		\label{opsumrep}
\end{equation}
is a completely positive map.  If in addition the $A_{\mu}$ operators 
satisfy
\begin{equation}
	\sum_{\mu} {A^{Q}_{\mu}}^{\dagger} A^{Q}_{\mu} = 1^{Q}
\end{equation}
then the map is also trace-preserving.  Such a representation for 
$\superop^{Q}$ in terms of operators $A^{Q}_{\mu}$ will be called
an ``operator-sum representation'' for $\superop^{Q}$.  
A single $\superop^{Q}$
will admit many different operator-sum representations.

Some insight into the connection between these representations 
for $\superop^{Q}$ can be gained by explicitly writing down the partial trace
$\tr_{E}$ from Equation \ref{bigunitary}.  
Suppose that $\rho^{Q} = \proj{\phi^{Q}}$ and let 
$\ket{\mu^{E}}$ be a complete orthonormal set of states of $E$.  Then
\begin{displaymath}
	\superop^{Q}(\rho^{Q}) = \sum_{\mu} \bra{\mu^{E}} U^{QE} 
		\left ( \proj{\phi^{Q}} \otimes \proj{0^{E}} \right )
		{U^{QE}}^{\dagger} \ket{\mu^{E}} .
\end{displaymath}
If we define the operator $A^{Q}_{\mu}$ by
\begin{displaymath}
	A^{Q}_{\mu} \ket{\phi^{Q}} = \bra{\mu^{E}} U^{QE} 
		\left ( \ket{\phi^{Q}} \otimes \ket{0^{E}} \right )
\end{displaymath}
then we recover an expression identical to Equation~\ref{opsumrep}.  Since
every input state $\rho^{Q}$ is a convex combination of pure states,
we recover Equation~\ref{opsumrep} for arbitrary $\rho^{Q}$ by linearity.

A pair of important representation theorems \cite{repthms} state
\begin{description}
	\item[I.]  Every trace-preserving, completely positive linear map 
		$\superop^{Q}$
		has a unitary representation, as in Equation~\ref{bigunitary}.
	\item[II.]  Every trace-preserving, completely positive linear map 
		$\superop^{Q}$ has an operator-sum representation, as in
		Equation~\ref{opsumrep}.
\end{description}
(By our argument above, the second statement follows from the first.)
These statements, particularly the first, motivate us to assert that the
trace-preserving, completely positive linear maps 
is exactly the class of allowed evolutions of a quantum system.  
Any reasonable evolution should be such
a map; and every such map could be accomplished by unitary dynamics 
(i.e., Hamiltonian evolution) on a larger system.

A relatively simple proof of both of these representation theorems is 
found in the Appendix.

From now on we will assume that a particular $\superop^{Q}$ has been 
specified, giving the evolution of states of the system $Q$.  We will
use unitary representations and operator-sum representations as convenient.

\section{Mixed States and Purifications}

\subsection{Entangled states}

Given a pure state $\ket{\Psi^{RQ}}$ of a joint system $RQ$, we can 
form the reduced state $\rho^{Q}$ for one of the subsystems $Q$ by means of
a partial trace operation:
\begin{eqnarray*}
	 \rho^{Q}   & = &   \tr_{R} \proj{\Psi^{RQ}}    \\
		& = &   \sum_{k} \amp{k^{R}}{\Psi^{RQ}} \amp{\Psi^{RQ}}{k^{R}}
\end{eqnarray*}
where $\ket{k^{R}}$ is an orthonormal basis for $\hilbert{R}$.  We can
define the reduced state $\rho^{Q}$ given a mixed joint state $\rho^{RQ}$
in the same fashion.

We have made use of a {\em partial inner product} between states of $R$ and 
states of a larger system $RQ$.  This is easy to understand.  The vector
\begin{displaymath}
	\ket{\xi^{Q}} = \amp{\phi^{R}}{\Psi^{RQ}}
\end{displaymath}
is defined to be the unique vector in $\hilbert{Q}$ such that
\begin{displaymath}
	\amp{\alpha^{Q}}{\xi^{Q}} = \amp{\phi^{R} \alpha^{Q}}{\Psi^{RQ}}
\end{displaymath}
for all vectors $\ket{\alpha^{Q}}$ in $\hilbert{Q}$ 
(where $\ket{\phi^{R} \alpha^{Q}} = \ket{\phi^{R}} \otimes \ket{\alpha^{Q}}$).
We could also write this as
\begin{displaymath}
	\amp{\phi^{R}}{\Psi^{RQ}} = \sum_{k} 
			\amp{\phi^{R} \xi^{Q}_{k}}{\Psi^{RQ}}
					\ket{\xi^{Q}_{k}}
\end{displaymath}
for some orthonormal basis set $\ket{\xi^{Q}_{k}}$ for $\hilbert{Q}$.

There are, of course, many different pure entangled states $\ket{\Psi^{RQ}}$
that give rise to a given reduced state $\rho^{Q}$.  These are generically
called {\em purifications} of $\rho^{Q}$.  Suppose $\ket{\Psi^{RQ}_{1}}$ and
$\ket{\Psi^{RQ}_{2}}$ are two such purifications.  Then we can write each
of them using the Schmidt decomposition:
\begin{eqnarray*}
	\ket{\Psi^{RQ}_{1}} & = &  \sum_{k} \sqrt{\lambda_{k}} 
			\ket{\xi^{R}_{1k}} \otimes \ket{\lambda^{Q}_{k}} \\
	\ket{\Psi^{RQ}_{2}} & = &  \sum_{k} \sqrt{\lambda_{k}} 
			\ket{\xi^{R}_{2k}} \otimes \ket{\lambda^{Q}_{k}}
\end{eqnarray*}
where the $\lambda_{k}$ and $\ket{\lambda^{Q}_{k}}$ are eigenvalues and
eigenstates of $\rho^{Q}$, and the $\ket{\xi^{R}_{1k}}$ and 
$\ket{\xi^{R}_{2k}}$ are two orthonormal sets of states in $\hilbert{R}$.  
Since the two purifications differ only in the choice of orthonormal set 
in $\hilbert{R}$,
they are connected by a unitary operator of the form $U^{R} \otimes 1^{Q}$.
Any purification of $\rho^{Q}$ can be converted to any other by
a unitary rotation acting on the auxilliary ``reference'' system $R$.

The Schmidt decomposition also makes clear the fact that, given a pure
entangled state 
\begin{displaymath}
	\ket{\Psi^{RQ}}  =   \sum_{k} \sqrt{\lambda_{k}} 
			\ket{\xi^{R}} \otimes \ket{\lambda^{Q}_{k}}
\end{displaymath}
the reduced states $\rho^{Q} = \tr_{R} \proj{\Psi^{RQ}}$ and 
$\rho^{R} = \tr_{Q} \proj{\Psi^{RQ}}$ will have exactly the same set
of non-zero eigenvalues, namely the $\lambda_{k}$.

\subsection{Mixed-state fidelity}

The notion of purification is used to define the {\em fidelity} between
two density operators $\rho_{1}$ and $\rho_{2}$.  This is
\begin{displaymath}
	F(\rho_{1},\rho_{2}) = \max \,\, \left | \amp{1}{2} \right |^{2}
\end{displaymath}
where the maximum is taken over all purifications $\ket{1}$ and $\ket{2}$ of
$\rho_{1}$ and $\rho_{2}$ \cite{qfidelity}.  
The fidelity has several important properties:
\begin{itemize}
	\item $0 \leq F(\rho_{1},\rho_{2}) \leq 1$, with $F(\rho_{1},\rho_{2})
		= 1$ if and only if $\rho_{1} = \rho_{2}$.
	\item $F(\rho_{1},\rho_{2}) = F(\rho_{2},\rho_{1})$.
	\item  If $\rho_{1} = \proj{\psi_{1}}$ is a pure state, then
		\begin{displaymath}
			F(\rho_{1},\rho_{2}) = \tr \, \rho_{1} \rho_{2} 
			= \bra{\psi_{1}} \rho_{2} \ket{\psi_{1}}.
		\end{displaymath}
		This is just the probability that the state $\rho_{2}$ 
		would pass a measurement testing whether or not it is
		the state $\ket{\psi_{1}}$.
\end{itemize}
The fidelity is a general way of defining the ``closeness'' of a pair of
states.

If we have two states $\rho^{RQ}_{1}$ and $\rho^{RQ}_{2}$, we can form
\begin{eqnarray*}
	\rho^{Q}_{1}    & = &   \tr_{R} \rho^{RQ}_{1}   \\
	\rho^{Q}_{2}    & = &   \tr_{R} \rho^{RQ}_{2}  .
\end{eqnarray*}
Then $F(\rho^{RQ}_{1},\rho^{RQ}_{2}) \leq F(\rho^{Q}_{1},\rho^{Q}_{2})$.
This can be seen directly from the definition by noting that every 
purification of $\rho^{RQ}_{1}$ is also a purification of $\rho^{Q}_{1}$,
and so on.

\subsection{Ensembles of pure states}

A mixed state $\rho^{Q}$ may arise from a statistical
ensemble ${\cal E}$ of pure states $\ket{\psi^{Q}_{i}}$ of $Q$.
In this case we can write
\begin{displaymath}
	\rho^{Q} = \sum_{i} p_{i} \proj{\psi^{Q}_{i}} ,
\end{displaymath}
where $p_{i}$ is the probability of the state $\ket{\psi^{Q}_{i}}$ in the
ensemble ${\cal E}$.

If $\rho^{Q} = \tr_{R} \proj{\Psi^{RQ}}$ for a pure entangled state
$\ket{\Psi^{RQ}}$ of $RQ$, we can ``realize'' an ensemble of pure states
for $\rho^{Q}$ by performing a complete measurement on the system $R$.
(This and other characterizations of the ensembles described 
by $\rho^{Q}$ are given in \cite{ensembles}.)
Let $\ket{\epsilon^{R}_{i}}$ be the basis for this complete measurement.
Each outcome of the $R$-measurement will be associated with a relative
state \cite{relstate}
of the system $Q$.  If $p_{i}$ is the probability of the $i$th 
outcome of the $R$-measurement and $\ket{\psi^{Q}_{i}}$ is the relative state 
of $Q$ associated with this outcome, then
\begin{displaymath}
	\sqrt{p_{i}} \, \ket{\psi^{Q}_{i}}  =  
		\amp{\epsilon^{R}_{i}}{\Psi^{RQ}} .
\end{displaymath}
(Note:  In dealing with ensembles of pure states, it is sometimes useful
to consider the non-normalized vectors $\ket{\tilde{\psi}^{Q}_{i}} = 
\sqrt{p_{i}} \, \ket{\psi^{Q}_{i}}$.  In other words, we can normalize
the component states in ${\cal E}$ by their probabilities.  The resulting
vectors are in themselves a complete description of the ensemble ${\cal E}$.
See \cite{ensembles} for fuller details.)
It follows that
\begin{eqnarray*}
\sum_{i} p_{i} \proj{\psi^{Q}_{i}} 
	& = & \sum_{i} \amp{\epsilon^{R}_{i}}{\Psi^{RQ}} 
		\amp{\Psi^{RQ}}{\epsilon^{R}_{i}} \\
	& = &   \tr_{R} \proj{\Psi^{RQ}} \\
	& = &   \rho^{Q} 
\end{eqnarray*}
so that the ensemble $\cal E$ of relative states is a pure state ensemble
for $\rho^{Q}$.  In fact, any pure state ensemble for $\rho^{Q}$
can be realized in just this way.  That is, we can fix a particular
purification $\ket{\Psi^{RQ}}$ for $\rho^{Q}$ and give a prescription for
realizing any pure state ensemble for $\rho^{Q}$ as a relative state
ensemble for some complete measurement on $R$.

Let ${\cal E}_{1}$ be a pure state ensemble for $\rho^{Q}$ given by 
probabilities $p_{i}$ and states $\ket{\psi^{Q}_{i}}$, and suppose that
$\hilbert{R}$ has arbitrarily high dimension, at least as large as 
the number of distinct pure states in the ensembles we consider.
Then we can construct a
purification $\ket{\Psi^{RQ}_{1}}$ by
\begin{displaymath}
	\ket{\Psi^{RQ}_{1}} = \sum_{i} \sqrt{p_{i}} \, 
		\ket{\alpha^{R}_{i}} \otimes \ket{\psi^{Q}_{i}}
\end{displaymath}
where the $\ket{\alpha^{R}_{i}}$ are a basis for $\hilbert{R}$.
(Only some of these basis vectors may appear in this superposition.)
Clearly, $\rho^{Q} = \tr_{R} \proj{\Psi^{RQ}_{1}}$.  Similarly, if we have
another ensemble ${\cal E}_{2}$ for $\rho^{Q}$ given by probabilities $q_{i}$
and states $\ket{\phi^{Q}_{i}}$, we can construct a purification
\begin{displaymath}
	\ket{\Psi^{RQ}_{2}} = \sum_{i} \sqrt{q_{i}} \, 
		\ket{\beta^{R}_{i}} \otimes \ket{\phi^{Q}_{i}}
\end{displaymath}
for some other $R$ basis $\ket{\beta^{R}_{i}}$.  Since both of these are
purifications of the same $\rho^{Q}$, there is a unitary operator $U^{R}$
such that $\ket{\Psi^{RQ}_{2}} = \left ( U^{R} \otimes 1^{Q} \right )
\ket{\Psi^{RQ}_{1}}$.  

We can clearly realize the ensemble ${\cal E}_{2}$
by making a measurement of the $\ket{\beta^{R}_{i}}$ basis on 
the state $\ket{\Psi^{RQ}_{2}}$ of $R$; but this
is equivalent to making a measurement of the basis
$\ket{\gamma^{R}_{i}} = {U^{R}}^{\dagger} \ket{\beta^{R}_{i}}$ 
on the state $\ket{\Psi^{RQ}_{1}}$:
\begin{eqnarray*}
	\amp{\gamma^{R}_{i}}{\Psi^{RQ}_{1}}
	& = & \left ( \bra{\beta^{R}_{i}} U^{R} \right ) 
		\ket{\Psi^{RQ}_{1}} \\
	& = & \bra{\beta^{R}_{i}} \left ( \left ( U^{R} \otimes 1^{Q} \right )
				\ket{\Psi^{RQ}_{1}} \right ) \\
	& = & \amp{\beta^{R}_{i}}{\Psi^{RQ}_{2}} \\
	& = & \sqrt{q_{i}} \ket{\phi^{Q}_{i}} .
\end{eqnarray*}
Thus, the ensemble ${\cal E}_{2}$ can be realized by making an $R$-measurement
on the purification $\ket{\Psi^{RQ}_{1}}$.

It follows that we could pick a particular purification $\ket{\Psi^{RQ}}$ 
and obtain {\em any} pure state ensemble for $\rho^{Q}$ by a suitable choice
of measurement basis for the system $R$.

We have assumed that $\dim \hilbert{R}$ is arbitrarily large so that
we can have an arbitrarily large number of basis vectors (since the pure
state ensembles may have an arbitrarily large number of components).  But
this is not really necessary.  If we allow positive operator measurements
(POMs) \cite{posops} on $R$, then the dimension of $\hilbert{R}$ need be no
greater than the dimension of $\hilbert{Q}$, which is the minimum size
necessary to purify all mixed states $\rho^{Q}$.  
The only relevant part of the basis $\ket{\alpha^{R}_{i}}$ is the set
of sub-normalized vectors $\ket{\tilde{\alpha}^{R}_{i}} = \Pi 
\ket{\alpha^{R}_{i}}$, where $\Pi$ is the projection onto the subspace
of $\hilbert{R}$ that supports $\rho^{R} = \tr_{Q} \proj{\Psi^{RQ}}$.
Since $\dim \hilbert{Q} = d$, this subspace need have only up to $d$
dimensions.  The $\proj{\tilde{\alpha}^{R}_{i}}$ are elements of a POM on 
this subspace.  We can use this POM on the $d$-dimensional subspace of
$\hilbert{R}$ to find a POM for a purification that uses another reference
system $R_{\ast}$, with $\dim \hilbert{R_{\ast}} = d$.

\subsection{Entropy}

Since entropy will be of central importance for our results, we will
review some of the relevant properties of classical and quantum entropy.

Suppose the non-negative numbers $p_{1}, p_{2}, \ldots$ 
sum to unity and thus form a probability distribution.  
The Shannon entropy 
$H(\vec{p})$ of this probability distribution (represented by the vector
$\vec{p}$) is just
\begin{equation}
	H(\vec{p}) = - \sum_{k} p_{k} \log p_{k} .
\end{equation}
We specify the base of our logarithms to be 2, and take $0 \log 0 = 0$. 
If $\vec{p}$ forms the probability for some random variable $X$,
so that $p(x_{k}) = p_{k}$ for various values 
$x_{k}$ of $X$, then we will often write this entropy as $H(X)$.

The Shannon entropy $H(X)$ is the fundamental quantity in classical information
theory, and it represents the average number of binary digits (or {\em bits})
required to represent the value of $X$ \cite{shannon}.  It can be thought
of as a measure of the uncertainty in the value of $X$ expressed by the
probability distribution.  We can use it to
define various information-theoretic quantities, such as the conditional
entropy
\begin{eqnarray*}
	H(X|Y)  & = &   \sum_{k} p(y_{k}) H(X|y_{k})    \\
		& = &   - \sum_{jk} p(x_{j},y_{k}) \log p(x_{j}|y_{k}) 
\end{eqnarray*}
for a joint distribution $p(x_{j},y_{k})$ 
over values of two variables $X$ and $Y$.
A very important quantity is the {\em mutual information} $I(X:Y)$ between
two random variables $X$ and $Y$: 
\begin{displaymath}
	I(X:Y) = H(X) - H(X|Y) ,
\end{displaymath}
which is the average amount that the uncertainty about $X$ decreases when
the value of $Y$ is known.  If $X$ represents the input of a communications
channel and $Y$ represents the output, then $I(X:Y)$ represents the amount
of information conveyed by the channel.  It turns out that $I(X:Y) = I(Y:X)$.

The quantum mechanical definition of entropy was first given by von Neumann
\cite{vonneumann}.  Suppose $\rho^{Q}$ is a density operator representing
a mixed state of $Q$.  Then the entropy is
\begin{equation}
	S(\rho^{Q}) = - \tr \, \rho^{Q} \log \rho^{Q} .
\end{equation}
If $\lambda_{1}, \lambda_{2}, \ldots$ are the eigenvalues of $\rho^{Q}$,
then $S(\rho^{Q}) = H(\vec{\lambda})$.  The von Neumann entropy 
also has a signficance for coding similar to the Shannon entropy: it is the
average number of two-level quantum systems (or {\em qubits}) needed to
faithfully represent one of the pure states of an ensemble described by
$\rho^{Q}$ \cite{qcoding}.

Suppose that systems $R$ and $Q$ are in a pure entangled state 
$\ket{\Psi^{RQ}}$.  Then $S(\rho^{RQ}) = 0$.  However, unlike the
classical Shannon entropy, it is possible for the von Neumann entropy
of the subsystems $R$ and $Q$ to be non-zero even when the entropy of 
the joint system $RQ$ is zero.  
We saw above that the density operators $\rho^{Q}$
and $\rho^{R}$ have the same non-zero eigenvalues.  Thus, $S(\rho^{R})
= S(\rho^{Q})$.  That is, if a pair of quantum systems are in a pure
entangled state, the reduced mixed states will have the same von Neumann
entropy.

The von Neumann entropy has a number of 
important properties (usefully reviewed in \cite{wehrl}).   
Suppose $A$ and $B$ are quantum systems
with joint state $\rho^{AB}$ and reduced states $\rho^{A}$ and $\rho^{B}$.
Then
\begin{eqnarray}
	S(\rho^{AB})  & \leq  & S(\rho^{A}) + S(\rho^{B})  \label{subadd} \\
	S(\rho^{AB})  & \geq  & S(\rho^{A}) - S(\rho^{B}) .  \label{triangle}
\end{eqnarray}
Equation~\ref{subadd} is the {\em subadditivity} property of the von
Neumann entropy, and Equation~\ref{triangle} is sometimes called
the ``triangle inequality'' for the entropy functional.

Another useful property of the von Neumann entropy relates it to the 
Shannon entropy of the probability distribution for the measurement 
outcomes of a complete observable.
Let $\rho$ be a mixed state with eigenvalues $\lambda_{k}$, so that
\begin{displaymath}
	\rho = \sum_{k} \lambda_{k} \proj{\lambda_{k}}.
\end{displaymath}
Now imagine that a measurement is performed of some complete ordinary
observable, that is, the state is resolved using an orthonormal basis 
$\ket{a_{j}}$.  The probability $p_{j}$ that the $j$th outcome
is obtained is thus
\begin{eqnarray*}
	p_{j}   & = &   \bra{a_{j}} \rho \ket{a_{j}}        \\
		& = &   \sum_{k} \lambda_{k} \amp{a_{j}}{\lambda_{k}}
				\amp{\lambda_{k}}{a_{j}}        \\
		& = &   \sum_{k} M_{jk} \lambda_{k} .
\end{eqnarray*}
The matrix $V_{jk} = \amp{a_{j}}{\lambda_{k}}$ is unitary, so the matrix
$M_{jk} = |V_{jk}|^{2}$ is doubly-stochastic.  That is, the rows and 
columns of $V_{jk}$ are orthonormal vectors, so that the rows and columns
of $M_{jk}$ all sum to one:
\begin{eqnarray*}
	\sum_{i} M_{ij} & = & 1 \mbox{  for all $j$} \\
	\sum_{j} M_{ij} & = & 1 \mbox{  for all $i$} .
\end{eqnarray*}
It is a standard theorem of information theory that the Shannon entropy
$H(\vec{q}) = -\sum_{i} q_{i} \log q_{i}$ cannot decrease if the probabilities
$q_{i}$ are changed via a doubly-stochastic matrix \cite{cover}.  Therefore, 
\begin{equation} 
	H(\vec{p}) \geq H(\vec{\lambda}) = S(\rho).  \label{entineq}
\end{equation}
The von Neumann entropy is thus a lower bound on the Shannon entropy for
the outcome of a complete measurement on the system.

\section{Entanglement fidelity}

\subsection{Definition}

Suppose that an entangled state $\ket{\Psi^{RQ}}$ is prepared for the
joint system $RQ$, and that $Q$ is subjected to a dynamical evolution
described by $\superop^{Q}$ (so that the overall evolution is given by
$\unity^{R} \otimes \superop^{Q}$).  The final state is 
\begin{displaymath}
\rho^{RQ'} = 
\unity^{R} \otimes \superop^{Q} \left( \proj{\Psi^{RQ}} \right) .
\end{displaymath}
The fidelity of this process is 
\begin{displaymath}
F_{e} = \tr \, \proj{\Psi^{RQ}} \rho^{RQ'} = \bra{\Psi^{RQ}} \rho^{RQ'}
					\ket{\Psi^{RQ}} .
\end{displaymath}
We call $F_{e}$ the {\em entanglement fidelity} of the process.

Written in these terms, $F_{e}$ depends on the initial and final states of
the system $RQ$.  We will next show that $F_{e}$ depends only on the
map $\superop^{Q}$ and the initial reduced state $\rho^{Q}$ obtained by
partial trace:
\begin{displaymath}
	\rho^{Q} = \tr_{R} \proj{\Psi^{RQ}} .
\end{displaymath}
That is, the entanglement fidelity $F_{e}$,
which is associated with an entangled state including $Q$, 
is (rather surprisingly) a property {\em intrinsic} to 
the system $Q$ itself.

The superoperator $\unity^{R} \otimes \superop^{Q}$ can be expressed
\begin{displaymath}
\unity^{R} \otimes \superop^{Q} \left ( \rho^{RQ}  \right )  =
	\sum_{\mu} \left ( 1^{R} \otimes A^{Q}_{\mu} \right )  \rho^{RQ}
	\left ( 1^{R} \otimes A^{Q}_{\mu} \right )^{\dagger} .
\end{displaymath}
Suppose that the initial states $\ket{\Psi^{RQ}_{1}}$ and 
$\ket{\Psi^{RQ}_{2}}$, both purifications of $\rho^{Q}$,
lead to final states $\rho^{RQ'}_{1}$ and $\rho^{RQ'}_{2}$, respectively,
under the action of the superoperator $\unity^{R} \otimes \superop^{Q}$;
and let $U^{R}$ be the unitary operator for $R$ such that
\begin{displaymath}
	\ket{\Psi^{RQ}_{2}}  =  
		\left ( U^{R} \otimes 1^{Q} \right ) \ket{\Psi^{RQ}_{1}} .
\end{displaymath}
Clearly, $U^{R} \otimes 1^{Q}$ commutes with $1^{R} \otimes A^{Q}_{\mu}$ for
all $\mu$.  Therefore,
\begin{eqnarray}
\rho^{RQ'}_{2} 
	& = &   \sum_{\mu} 
		\left ( 1^{R} \otimes A^{Q}_{\mu} \right ) \proj{\Psi^{RQ}_{2}}
		\left ( 1^{R} \otimes A^{Q}_{\mu} \right )^{\dagger}  
		\nonumber \\
	& = &   \sum_{\mu}
		\left ( 1^{R} \otimes A^{Q}_{\mu} \right )
		\left ( U^{R} \otimes 1^{Q}  \right ) \proj{\Psi^{RQ}_{1}}
		\left ( U^{R} \otimes 1^{Q}  \right )^{\dagger}
		\left ( 1^{R} \otimes A^{Q}_{\mu} \right ) \nonumber \\
	& = &   \left ( U^{R} \otimes 1^{Q} \right) 
		\left ( \sum_{\mu}
		\left ( 1^{R} \otimes A^{Q}_{\mu} \right ) \proj{\Psi^{RQ}_{1}}
		\left ( 1^{R} \otimes A^{Q}_{\mu} \right )^{\dagger} \right )
		\left ( U^{R} \otimes 1^{Q} \right )^{\dagger} \nonumber \\
\rho^{RQ'}_{2}
	& = &   \left ( U^{R} \otimes 1^{Q} \right ) \rho^{RQ'}_{1}
		\left ( U^{R} \otimes 1^{Q} \right )^{\dagger} . 
		\label{unrel}
\end{eqnarray}
(Note that equation \ref{unrel} 
implies that $\rho^{RQ'}_{1}$ and
$\rho^{RQ'}_{2}$ must have the same eigenvalues.  This will
be important later in the definition of entropy production.)
From equation \ref{unrel} it follows that 
\begin{eqnarray*}
F_{e2}  & = &   \bra{\Psi^{RQ}_{2}} \rho^{RQ'}_{2} \ket{\Psi^{RQ}_{2}} \\
	& = &   \bra{\Psi^{RQ}_{1}} 
		\left ( U^{R} \otimes 1^{Q} \right )^{\dagger} 
		\left ( U^{R} \otimes 1^{Q} \right ) 
		\rho^{RQ'}_{1}  
		\left ( U^{R} \otimes 1^{Q} \right )^{\dagger} 
		\left ( U^{R} \otimes 1^{Q} \right ) 
		\ket{\Psi^{RQ}_{1}} \\
	& = &   \bra{\Psi^{RQ}_{1}} \rho^{RQ'}_{1} \ket{\Psi^{RQ}_{1}} \\
	& = &   F_{e1} .
\end{eqnarray*}
Hence, the fidelity $F_{e}$ does not depend on {\em which} purification for
$\rho^{Q}$ is chosen.  It only depends on $\rho^{Q}$ and the superoperator
$\superop^{Q}$.

\subsection{Intrinsic expression for $F_{e}$}

It is instructive to derive an expression for $F_{e}$ in terms of things
that are intrinsic to the system $Q$---i.e., an expression that
does not refer to $R$.  Suppose we 
have an operator-sum representation for $\superop^{Q}$, as in 
Equation~\ref{opsumrep}.  Consider a particular pure entangled
state for $RQ$
\begin{displaymath}
\ket{\Psi^{RQ}} = \sum_{k} \sqrt{p_{k}} \ket{k^{R}} \otimes \ket{\phi^{Q}_{k}}
\end{displaymath}
where the $\ket{k^{R}}$ are orthonormal states in $\hilbert{R}$.  
(We do not need to require
the $\ket{\phi^{Q}_{k}}$ to be orthonormal.)
This state evolves under $\unity^{R} \otimes \superop^{Q}$ into $\rho^{RQ'}$.
The initial state of $Q$ is 
\begin{displaymath}
   \rho^{Q} = \tr_{R} \proj{\Psi^{RQ}} = \sum_{k} p_{k} \proj{\phi^{Q}_{k}} .
\end{displaymath}
Now, for any operator $X^{Q}$ acting on $\hilbert{Q}$,
\begin{eqnarray*}
\bra{\Psi^{RQ}} \left ( 1^{R} \otimes X^{Q} \right ) \ket{\Psi^{RQ}}
	& = & \sum_{jk} \sqrt{p_{j} p_{k}} 
		\bra{j^{R}} 1^{R} \ket{k^{R}}
		\bra{\phi^{Q}_{j}} X^{Q} \ket{\phi^{Q}_{k}} \\
	& = & \sum_{jk} \sqrt{p_{j} p_{k}} \delta_{jk}
		\bra{\phi^{Q}_{j}} X^{Q} \ket{\phi^{Q}_{k}} \\
	& = & \sum_{k} p_{k} \, \bra{\phi^{Q}_{k}} X^{Q} \ket{\phi^{Q}_{k}} \\
	& = & \tr \, \rho^{Q} X^{Q} .
\end{eqnarray*}
We can now work out the fidelity very easily:
\begin{eqnarray}
F_{e}   & = &   \bra{\Psi^{RQ}} \rho^{RQ'} \ket{\Psi^{RQ}}  \nonumber \\
	& = &   \sum_{\mu} \bra{\Psi^{RQ}} 
		\left ( 1^{R} \otimes A^{Q}_{\mu} \right )
		\proj{\Psi^{RQ}}
		\left ( 1^{R} \otimes A^{Q}_{\mu} \right )^{\dagger}
		\ket{\Psi^{RQ}}                 \nonumber \\
F_{e}   & = &   \sum_{\mu}      
			\left ( \tr \, \rho^{Q} A^{Q}_{\mu} \right )
			\left ( \tr \, \rho^{Q} {A^{Q}_{\mu}}^{\dagger} \right ) .
			\label{fedef}
\end{eqnarray}
Although this is written with respect to a particular operator-sum
representation of $\superop^{Q}$ (which is not unique), the value of $F_{e}$ 
will clearly be independent of this representation.  Equation~\ref{fedef} 
expresses $F_{e}$ entirely in terms of the initial state $\rho^{Q}$ of
the system $Q$ and the evolution superoperator $\superop^{Q}$.

\subsection{Relations to other fidelities}

It is worth noting what $F_{e}$ is not.  It is not the simple fidelity
of the input and output states of $Q$.  This fidelity can be written
$F(\rho^{Q},\rho^{Q'})$, where $\rho^{Q'} = \superop^{Q} ( \rho^{Q} )$.
We can show that $F_{e} \neq F(\rho^{Q},\rho^{Q'})$ in general by 
considering an operation defined by
\begin{displaymath}
	A^{Q}_{\mu} = \proj{\mu^{Q}}
\end{displaymath}
for some orthonormal basis $\ket{\mu^{Q}}$.  The effect of the operation 
is to completely destroy any coherences between different elements of the
basis.  That is, the superposition $\sum_{\mu} c_{\mu} \ket{\mu^{Q}}$ 
would be transformed into the mixed state 
\begin{displaymath}
	\rho^{Q'} = \sum_{\mu} |c_{\mu}|^{2} \proj{\mu^{Q}} .
\end{displaymath}
Now suppose $\rho^{Q} = \sum_{\mu} \lambda_{\mu} \proj{\mu^{Q}}$.  Then
$\rho^{Q'} = \rho^{Q}$ and thus $F(\rho^{Q},\rho^{Q'}) = 1$.  However,
let $\ket{\Psi^{RQ}}$ be a purification of $\rho^{Q}$, for example
\begin{displaymath}
	\ket{\Psi^{RQ}} = \sum_{\mu} \sqrt{\lambda_{\mu}} \ket{\phi^{R}_{\mu}}
					\otimes \ket{\mu^{Q}} .
\end{displaymath}
The action of the superoperator $\unity^{R} \otimes \superop^{Q}$ on this
state yields
\begin{displaymath}
	\rho^{RQ'} = \sum_{\mu} \lambda_{\mu} \proj{\phi^{R}_{\mu}} \otimes
						\proj{\mu^{Q}} .
\end{displaymath}
If more than one of the $\lambda_{\mu}$'s is non-zero, then $F_{e} 
= F(\rho^{RQ},\rho^{RQ'}) \neq 1$.
Thus, $F_{e} \neq F(\rho^{Q},\rho^{Q'})$.

However, there is a general relation between 
$F_{e}$ and $F(\rho^{Q},\rho^{Q'})$.
\begin{equation}
	F_{e} = F(\rho^{RQ},\rho^{RQ'}) \leq F(\rho^{Q},\rho^{Q'}) .
\end{equation}
The entanglement fidelity $F_{e}$ is thus a lower bound to the ``input-output''
fidelity $F(\rho^{Q},\rho^{Q'})$ for states of $Q$.

$F_{e}$ and $F(\rho^{Q},\rho^{Q'})$ do sometimes agree.
Suppose that the initial state $\rho^{Q}$ is in fact a pure state of $Q$,
so that there is no entanglement between $R$ and $Q$.  Then, letting
$\rho^{Q} = \proj{\psi^{Q}}$,
\begin{eqnarray*}
	F(\rho^{Q},\rho^{Q'})
		& = &   \bra{\psi^{Q}} \rho^{Q'} \ket{\psi^{Q}}     \\
		& = &   \sum_{\mu} \bra{\psi^{Q}} A^{Q}_{\mu} 
			\proj{\psi^{Q}} {A^{Q}_{\mu}}^{\dagger} 
			\ket{\psi^{Q}}                  \\
		& = &   \sum_{\mu} \left ( \tr \, \rho^{Q} A^{Q}_{\mu} \right )
			\left (\tr \, \rho^{Q} {A^{Q}_{\mu}}^{\dagger} \right ) \\
		& = &   F_{e} .
\end{eqnarray*}
The entanglement fidelity equals the ``input-output'' fidelity when the 
input state is a pure state.

Now suppose that $\rho^{Q}$ is a mixed state of $Q$ arising from an 
ensemble ${\cal E}$ in which the pure state $\ket{\psi^{Q}_{i}}$ occurs
with probability $p_{i}$.  The average ``input-output'' fidelity for
this ensemble is
\begin{eqnarray*}
  \bar{F} & = & \sum_{i} p_{i} F \left ( \proj{\psi^{Q}_{i}},
				\rho^{Q'}_{i} \right ) \\
	  & = & \sum_{i} p_{i} \bra{\psi^{Q}_{i}} \rho^{Q'}_{i} 
				\ket{\psi^{Q}_{i}}
\end{eqnarray*}
where $\rho^{Q'}_{i} = \superop^{Q}(\proj{\psi^{Q}_{i}})$.

It turns out that $\bar{F} \geq F_{e}$.  Some such connection is
reasonable physically,
since we can ``realize'' a pure state ensemble ${\cal E}$ by means of
an $R$-measurement on a purification of $\rho^{Q}$, and this measurement
may be performed either before or after the dynamical evolution given 
by $\superop^{Q}$.  A full proof follows:

Let $\ket{\alpha^{R}_{i}}$ be
an orthonormal set in $\hilbert{R}$ (assumed to have as many dimensions
as there are elements in the ensemble ${\cal E}$), and let
\begin{displaymath}
	\ket{\Psi^{RQ}} = \sum_{i} \sqrt{p_{i}} \ket{\alpha^{R}_{i}}
				\otimes \ket{\psi^{Q}_{i}} .
\end{displaymath}
$\ket{\Psi^{RQ}}$ is clearly a purification of $\rho^{Q}$, and the 
$\ket{\alpha^{R}_{i}}$ basis is the basis in $\hilbert{R}$ which,
when measured, generates the ensemble  ${\cal E}$ as an ensemble 
of relative states in $Q$.  That is,
$\sqrt{p_{i}} \ket{\psi^{Q}_{i}} = \amp{\alpha^{R}_{i}}{\Psi^{RQ}}$,
which we could also write as
\begin{displaymath}
	\left ( \proj{\alpha^{R}_{i}} \otimes 1^{Q} \right )
		\ket{\Psi^{RQ}}  =  \sqrt{p_{i}} \ket{\alpha^{R}_{i}}
					\otimes \ket{\psi^{Q}_{i}} .
\end{displaymath}
Now consider the operator $\Gamma^{RQ}$ given by
\begin{eqnarray*}
\Gamma^{RQ}
	& = &   \sum_{j} \proj{\alpha^{R}_{j}} \otimes \proj{\psi^{Q}_{j}} \\
	& = &   \sum_{j} \left ( 1^{R} \otimes \proj{\psi^{Q}_{j}} \right )
			\left ( \proj{\alpha^{R}_{j}} \otimes 1^{Q} \right ) .
\end{eqnarray*}
Since $\Gamma^{RQ}$ is the sum of an orthogonal set of projections, it is 
itself a projection operator onto some subspace of $\hilbert{R} \otimes
\hilbert{Q}$.  $\ket{\Psi^{RQ}}$ itself is in this subspace:
\begin{eqnarray*}
\Gamma^{RQ} \ket{\Psi^{RQ}}
	& = &   \sum_{j} \left ( 1^{R} \otimes \proj{\psi^{Q}_{j}} \right )
			\left ( \proj{\alpha^{R}_{j}} \otimes 1^{Q} \right )
			\ket{\Psi^{RQ}}     \\
	& = &   \sum_{j} \left ( 1^{R} \otimes \proj{\psi^{Q}_{j}} \right )
			\sqrt{p_{j}} \ket{\alpha^{R}_{j}} 
				\otimes \ket{\psi^{Q}_{j}}  \\
	& = &   \sum_{j} \sqrt{p_{j}} \ket{\alpha^{R}_{j}} 
				\otimes \ket{\psi^{Q}_{j}}  \\
	& = &   \ket{\Psi^{RQ}} .
\end{eqnarray*}
Therefore, we have the operator inequality $\Gamma^{RQ} \geq \proj{\Psi^{RQ}}$.
This means that, for any vector $\ket{\chi^{RQ}}$,
\begin{displaymath}
	\bra{\chi^{RQ}} \Gamma^{RQ} \ket{\chi^{RQ}} \geq 
		\bra{\chi^{RQ}} \left ( \proj{\Psi^{RQ}} \right )
		\ket{\chi^{RQ}} 
\end{displaymath}
which in turn implies that, for all positive operators $X^{RQ}$,
\begin{displaymath}
	\tr \, \Gamma^{RQ} X^{RQ} \geq \tr \, \proj{\Psi^{RQ}} X^{RQ}
				= \bra{\Psi^{RQ}} X^{RQ} \ket{\Psi^{RQ}} .
\end{displaymath}

Let $A^{Q}_{\mu}$ be the operators in an operator-sum representation of the 
evolution superoperator $\superop^{Q}$.  Then
\begin{eqnarray*}
\lefteqn{\Gamma^{RQ} 
\left ( 1^{R} \otimes A^{Q}_{\mu} \right ) \ket{\Psi^{RQ}}} \\
	& = &   \sum_{j} \left ( 1^{R} \otimes \proj{\psi^{Q}_{j}} \right )
			\left ( \proj{\alpha^{R}_{j}} \otimes 1^{Q} \right )
			\left ( 1^{R} \otimes A^{Q}_{\mu} \right )
			\ket{\Psi^{RQ}}     \\
	& = &   \sum_{j} \left ( 1^{R} \otimes \proj{\psi^{Q}_{j}} \right )
			\left ( 1^{R} \otimes A^{Q}_{\mu} \right )
			\left ( \proj{\alpha^{R}_{j}} \otimes 1^{Q} \right )
			\ket{\Psi^{RQ}}     \\
	& = &   \sum_{j} \left ( 1^{R} \otimes \proj{\psi^{Q}_{j}} \right )
			\left ( 1^{R} \otimes A^{Q}_{\mu} \right )
			\sqrt{p_{j}} \ket{\alpha^{R}_{j}} \otimes
					\ket{\psi^{Q}_{j}}  \\
	& = &   \sum_{j} \left ( 1^{R} \otimes \proj{\psi^{Q}_{j}} \right )
			\sqrt{p_{j}} \ket{\alpha^{R}_{j}} \otimes
					A^{Q}_{\mu} \ket{\psi^{Q}_{j}} \\
	& = &   \sum_{j} \sqrt{p_{j}} \bra{\psi^{Q}_{j}} A^{Q}_{\mu}
			\ket{\psi^{Q}_{j}}  \ket{\alpha^{R}_{j}}
			\otimes \ket{\psi^{Q}_{j}} .
\end{eqnarray*}
If $\rho^{RQ'} = \unity^{R} \otimes \superop^{Q} \left (
	\proj{\Psi^{RQ}} \right )$, then
\begin{eqnarray*}
F_{e}   & = & \tr \, \proj{\Psi^{RQ}} \rho^{RQ'}   \\
	& \leq & \tr \, \Gamma^{RQ} \rho^{RQ'} \\
	& = &   \tr \, \Gamma^{RQ} \rho^{RQ'} \Gamma^{RQ}  \\
	& = &   \sum_{\mu} \tr \, \Gamma^{RQ} 
		\left ( 1^{R} \otimes A^{Q}_{\mu} \right ) \proj{\Psi^{RQ}}
		\left ( 1^{R} \otimes A^{Q}_{\mu} \right )^{\dagger}
		\Gamma^{RQ}  \\
	& = &   \sum_{jk} \sum_{\mu} \sqrt{p_{j} p_{k}} 
		\bra{\psi^{Q}_{j}} A^{Q}_{\mu} \ket{\psi^{Q}_{j}}
		\bra{\psi^{Q}_{k}} {A^{Q}_{\mu}}^{\dagger} 
		\ket{\psi^{Q}_{k}}
		\amp{\psi^{Q}_{k}}{\psi^{Q}_{j}}
		\amp{\alpha^{R}_{k}}{\alpha^{R}_{j}}    \\
	& = &   \sum_{k} \sum_{\mu} p_{k} 
		\bra{\psi^{Q}_{k}} A^{Q}_{\mu} \ket{\psi^{Q}_{k}}
		\bra{\psi^{Q}_{k}} {A^{Q}_{\mu}}^{\dagger} 
		\ket{\psi^{Q}_{k}}  \\
	& = &   \sum_{k} p_{k} \bra{\psi^{Q}_{k}}
		\left ( \sum_{\mu} A^{Q}_{\mu} \proj{\psi^{Q}_{k}}
		{A^{Q}_{\mu}}^{\dagger} \right )  \ket{\psi^{Q}_{k}}    \\
	& = &   \sum_{k} p_{k} \bra{\psi^{Q}_{k}} \rho^{Q'}_{k}
			\ket{\psi^{Q}_{k}}  \\
	& = &   \bar{F} .
\end{eqnarray*}
Thus, $\bar{F} \geq F_{e}$, as we wished to show.  
The average ``input-output''
fidelity under the evolution superoperator $\superop^{Q}$ for any ensemble
of pure states with density operator $\rho^{Q}$ is bounded below by the
entanglement fidelity $F_{e}$.

\section{Entropy production}

\subsection{Definition}

As was shown in Equation~\ref{unrel} above, if $\ket{\Psi^{RQ}_{1}}$ and
$\ket{\Psi^{RQ}_{2}}$ are two purifications of $\rho^{Q}$, and each is
subjected to the same evolution superoperator 
$\unity^{R} \otimes \superop^{Q}$, the resulting states $\rho^{RQ'}_{1}$
and $\rho^{RQ'}_{2}$ will have exactly the same eigenvalues.  Therefore,
\begin{displaymath}
	S(\rho^{RQ'}_{1}) = S(\rho^{RQ'}_{2})
\end{displaymath}
where $S(\rho)$ is the von Neumann entropy of the
density operator $\rho$.  In other words, the entropy of the final joint
state of $RQ$ is independent of which purification is chosen.  
Again, rather surprisingly, we have a quantity that depends only on
the initial state $\rho^{Q}$ and the evolution superoperator $\superop^{Q}$;
that is, we have a quantity that is {\em intrinsic} to $Q$.

For a given $\rho^{Q}$ and $\superop^{Q}$, we therefore define the 
{\em entropy production} $S_{e}$ to be
\begin{equation}
	S_{e} = - \tr \, \rho^{RQ'} \log \rho^{RQ'}
\end{equation}
where $\rho^{RQ'} = \unity^{R} \otimes \superop^{Q} \left ( \proj{\Psi^{RQ}}
\right )$ and $\ket{\Psi^{RQ}}$ is some purification of $\rho^{Q}$.

We will now derive an explicit expression for $S_{e}$ in terms of 
$\rho^{Q}$ and $\superop^{Q}$.  
Suppose we have an operator-sum representation for $\superop^{Q}$,
and we define
\begin{displaymath}
	\ket{\tilde{\Phi}^{RQ'}_{\mu}} = \left ( 1^{R} \otimes A^{Q}_{\mu}
				\right ) \ket{\Psi^{RQ}} .
\end{displaymath}
(These are not normalized vectors in general.)  Then
\begin{eqnarray*}
	\rho^{RQ'} & = & \sum_{\mu} \left ( 1^{R} \otimes A^{Q}_{\mu} \right )
			\proj{\Psi^{RQ}}
		\left ( 1^{R} \otimes A^{Q}_{\mu} \right )^{\dagger} \\
		& = & \sum_{\mu} \proj{\tilde{\Phi}^{RQ'}_{\mu}} .
\end{eqnarray*}
Thus, the vectors $\ket{\tilde{\Phi}^{RQ'}_{\mu}}$ give us a pure state
ensemble for $\rho^{RQ'}$.
We can use these states to construct a purification for $\rho^{RQ'}$.  
Let us adjoin a system $E$ whose Hilbert space 
$\hilbert{E}$ has at least as many
dimensions as the number of $A^{Q}_{\mu}$ operators.  Then the state
\begin{displaymath}
	\ket{\Upsilon^{RQE'}} = \sum_{\mu} \ket{\tilde{\Phi}^{RQ'}_{\mu}}
				\otimes \ket{\mu^{E}}
\end{displaymath}
(where the $\ket{\mu^{E}}$ are an orthonormal set of $E$ states) will be
a purification for $\rho^{RQ'}$.

Since the state $\ket{\Upsilon^{RQE'}}$ is a pure state, the reduced states
\begin{eqnarray*}
	\rho^{RQ'} & = & \tr_{E} \proj{\Upsilon^{RQE'}} \\
	\rho^{E'} & = & \tr_{E} \proj{\Upsilon^{RQE'}}
\end{eqnarray*}
will have exactly the same non-zero
eigenvalues.  Therefore, $S_{e} = S(\rho^{RQ'}) = S(\rho^{E'})$.  We
can write down the density operator $\rho^{E'}$:
\begin{eqnarray*}
\rho^{E'}   & = &   \tr_{RQ} \proj{\Upsilon^{RQE'}} \\
		& = &   \sum_{\mu,\nu} 
		\amp{\tilde{\Phi}^{RQ'}_{\nu}}{\tilde{\Phi}^{RQ'}_{\mu}}
			\ket{\mu^{E}} \bra{\nu^{E}} .
\end{eqnarray*}
That is, $\rho^{E'} = \sum_{\mu \nu} W_{\mu \nu} \ket{\mu^{E}} \bra{\nu^{E}}$,
where
\begin{eqnarray*}
  W_{\mu \nu}   & = & \amp{\tilde{\Phi}^{RQ'}_{\nu}
			}{\tilde{\Phi}^{RQ'}_{\mu}} \\
		& = & \tr_{RQ} \ket{\tilde{\Phi}^{RQ'}_{\mu}}
				\bra{\tilde{\Phi}^{RQ'}_{\nu}} \\
		& = & \tr_{RQ} \left ( 1^{R} \otimes A^{Q}_{\mu} \right )
			\proj{\Psi^{RQ'}}
			\left ( 1^{R} \otimes 
			A^{Q}_{\nu} \right )^{\dagger} \\
		& = & \tr_{Q}  A^{Q}_{\mu}
			\left ( \tr_{R} \proj{\Psi^{RQ}} \right )
			{A^{Q}_{\nu}}^{\dagger}  \\
		& = & \tr_{Q}  A^{Q}_{\mu} \rho^{Q} {A^{Q}_{\nu}}^{\dagger} .
\end{eqnarray*}
In other words we have the following prescription.  Let $W$ be a density
operator with components (in some orthonormal basis) 
\begin{equation}
	W_{\mu \nu} = \tr \,  A^{Q}_{\mu} \rho^{Q} {A^{Q}_{\nu}}^{\dagger}.
		\label{wopdef}
\end{equation}
Then
\begin{equation}
	S_{e} = S(W) .
\end{equation}

As explained in the Appendix, any two operator-sum representations for
$\superop^{Q}$ are related by a unitary matrix $U_{\mu \nu}$.  This
simply corresponds to the freedom to write the matrix $W_{\mu \nu}$
with respect to any basis (which obviously does not affect $S_{e}$).  
Let $P_{\mu} = W_{\mu \mu}$ be the diagonal elements of $W_{\mu \nu}$.
These would be the probabilities given the state $W$
for a complete measurement using the basis that yields the
matrix elements $W_{\mu \nu}$.  Therefore, $H(\vec{P}) \geq S(W)$.
But we could, by choosing the unitary matrix that diagonalizes
$W_{\mu \nu}$, find a representation such that $H(\vec{P}) = S(W)$.
This yields another expression for $S_{e}$:
\begin{equation}
	S_{e} = \min \left ( - \sum_{\mu} P_{\mu} \log P_{\mu} \right )
\end{equation}
where $P_{\mu} = \tr \, A^{Q}_{\mu} \rho^{Q} {A^{Q}_{\mu}}^{\dagger}$ and
the minimum is taken over all operator-sum representations of 
$\superop^{Q}$.

For a given input state $\rho^{Q}$, there is a ``diagonal'' operator-sum 
representation, in which $W_{\mu \nu}$ is diagonal.  In this representation,
\begin{displaymath}
	\tr \, A^{Q}_{\mu} \rho^{Q} {A^{Q}_{\nu}}^{\dagger} 
			=  0  \mbox{    for $\mu \neq \nu$}  .
\end{displaymath}
If $\rho^{Q} = d^{-1} 1^{Q}$ (the ``maximally mixed'' state), then
this simply means that the various $A^{Q}_{\mu}$ operators are 
orthogonal in the operator inner product $\langle B , C \rangle =
\tr \, B^{\dagger} C$.
This diagonal representation is minimal, in the sense that no other
operator-sum representation includes a smaller number of $A^{Q}_{\mu}$
operators.

The evolution $\superop^{Q}$ might in fact be due to unitary evolution of
a larger system that includes an environment $E$, with $E$ initially in
a pure state and $RQ$ initially in a pure entangled state.
In this case the final state of $RQE$ will be also be a pure state.
Then $S(\rho^{E'}) = S(\rho^{RQ'}) = S_{e}$.
In other words, the
entropy production $S_{e}$is just the entropy produced in the environment,
if it is initially in a pure state.

Note that the same $\rho^{E'}$ would have been obtained if we ignored
the reference system $R$ entirely and simply considered the unitary
evolution of $QE$ with an initial state $\rho^{Q}$ for $Q$.  The
entropy produced in the environment does not depend on the dynamically
isolated reference system $R$.

The assumption that the environment is initially in a pure state 
$\ket{0^{E}}$ at first seems too restrictive.  For example, we may wish
to consider environments that are initially in some thermal equilibrium
state $\rho^{E}$.  However, we may imagine that the environment
consists of a ``near'' environment $E_{n}$ and a ``far'' environment
$E_{f}$.  The system $Q$ only interacts with the near environment $E_{n}$.
The initial state of the full environment may be an entangled pure state, 
but the system $Q$ will ``see'' a mixed state for $E_{n}$.

To summarize, the entropy production $S_{e}$ has the following properties:
\begin{itemize}
	\item  $S_{e}$ is a quantity intrinsic to the system $Q$, and
		can be defined entirely in terms of the initial state
		$\rho^{Q}$ and the superoperator $\superop^{Q}$.
	\item  If the initial state $\rho^{Q}$ arises because a larger
		system $RQ$ is in a pure entangled state, and if the
		reference system $R$ has trivial dynamics, then the
		entropy production $S_{e}$ is the entropy of the final
		state $\rho^{RQ'}$ of $RQ$.  (It is easy to generalize
		this to the case when $R$ itself can have arbitrary unitary 
		evolution---i.e., when $R$ is dynamically isolated
		but may have a non-zero internal Hamiltonian.)
	\item  If the non-unitary evolution of $Q$ arises because $Q$ 
		interacts with an environment $E$ that is initially in 
		a pure state, then $S_{e}$ is the entropy of the final
		state $\rho^{E'}$ of the environment.
	\item  If the initial state $\rho^{Q}$ of the system $Q$ is
		a pure state, we can adopt a unitary representation
		for $\superop^{Q}$ in which $E$ is also initially
		in a pure state.  Then $\rho^{Q'}$ and $\rho^{E'}$
		have the same eigenvalues.  In this case, $S_{e} =
		S(\rho^{Q'})$, the entropy produced in the system $Q$.
\end{itemize}

\subsection{Relation to other entropies}

Once again, it is useful to emphasize what $S_{e}$ is not.  It is not
in general the increase in the entropy of the system $Q$---in fact,
this entropy may actually decrease, wheras $S_{e}$ is never negative.
It is also not always the entropy increase of the environment, if the
initial environment state is mixed.  The entropy production $S_{e}$ 
characterizes the {\em information exchange} between the system $Q$
and the external world during the evolution given by $\superop^{Q}$.

There are, however, inequalities relating $S_{e}$ to entropy changes
in $Q$ and $E$.  
First we will relate the entropy production to changes in the 
entropy of $Q$.  Suppose an evolution superoperator $\superop^{Q}$
is given, together with an initial state $\rho^{Q}$ of $Q$.  
We can always find a representation for $\superop^{Q}$
as a unitary evolution on a larger system $QE$ with an initial
pure state $\ket{0^{E}}$ for the environment system.  With this
representation, the
entropy of the joint initial state $S(\rho^{QE}) = S(\rho^{Q})$.  The
joint system $QE$ evolves unitarily, so the entropy of the joint state
remains unchanged.  Thus, $S(\rho^{QE'}) = S(\rho^{Q})$.  The entropy
production in this case is the final entropy of the environment 
$S(\rho^{E'})$.  The triangle inequality (equation \ref{triangle}) 
yields
\begin{eqnarray}
	S(\rho^{Q}) & \geq & S(\rho^{Q'}) - S(\rho^{E'}) \nonumber \\
	S_{e} & \geq & S(\rho^{Q'}) - S(\rho^{Q}) .
\end{eqnarray}
In other words, the entropy production is no less than the increase in
entropy of the system $Q$.  We can also in this way establish that
\begin{equation}
	S_{e}  \leq  S(\rho^{Q}) + S(\rho^{Q'}) .
\end{equation}

Now we relate $S_{e}$ to the entropy change in the environment.  In
this case, we are given a particular (possibly mixed) initial state
$\rho^{E}$ for the environment and a particular unitary evolution
$U^{QE}$ for the joint system $QE$.  Again, the initial state of $Q$
is $\rho^{Q}$, but now we will imagine that this is a partial state
of a pure entangled state $\ket{\Psi^{RQ}}$, where $R$ is an isolated
reference system.  The entropy of the joint system $RQE$ is initially
$S(\rho^{RQE}) = S(\rho^{E})$, and remains unchanged during 
the unitary evolution of the joint system.  By definition, the entropy
production is just the entropy $S(\rho^{RQ'})$ of the final state of $RQ$.
Thus,
\begin{eqnarray}
	S(\rho^{E}) & \geq & S(\rho^{E'}) - S(\rho^{RQ'}) \nonumber \\
	S_{e} & \geq & S(\rho^{E'}) - S(\rho^{E}) ,
\end{eqnarray}
so that the entropy production is no less than the increase in the entropy
of the environment.  We can also derive
\begin{equation}
	S_{e} \leq S(\rho^{E}) + S(\rho^{E'}) ,
\end{equation}
which, for a large environment, is probably not very useful.

Similar arguments based on the subaddtivity of the entropy functional,
Equation~\ref{subadd},
also demonstrate that $S_{e}$ is no smaller than the entropy
{\em decrease} in either the system $Q$ or the environment $E$.
To summarize the lower bounds for $S_{e}$,
\begin{eqnarray*}
	S_{e} & \geq & \left | \Delta S^{Q} \right |  \\
	S_{e} & \geq & \left | \Delta S^{E} \right |
\end{eqnarray*}
where $\Delta S^{Q}$ and $\Delta S^{E}$ are the changes in entropy of
the system $Q$ and environment $E$, respectively.

\subsection{Entropy production and eavesdropping}

There is a simple application of these ideas to quantum cryptography
\cite{qcrypto}.
Suppose Alice prepares the state $\rho^{Q}_{k}$ of $Q$ with probability
$p_{k}$, and then conveys the system $Q$ to Bob as part of a quantum
cryptographic protocol.  (Alternatively, we could imagine that Alice
prepares $Q$ in a state entangled with a system $R$, which she retains,
as part of an entanglement-based protocol \cite{qcrypto-entang}.  
But in such protocols,
Alice usually later makes a measurement on $R$, giving rise to an
ensemble of relative states of $Q$.)
Along the way $Q$ may interact with the rest of
the world, represented by the environment system $E$, producing some
level of ``noise'' in $Q$.  The environment, however, may also contain
the measuring apparatus of an eavesdropper Eve.  We will assume that
the environment is initially in a pure state (but see the remark
above about the possibility of an entangled state of ``near'' and 
``far'' zones within the environment).

The dynamical evolution of $Q$ is given by the evolution superoperator
$\superop^{Q}$.  Let $S_{e,k}$ be the entropy production in $Q$ for 
the input state $\rho^{Q}_{k}$, which equals the entropy of the final
environment state $\rho^{E'}_{k}$ resulting from the input of 
$\rho^{Q}_{k}$; and let $S_{e}$ be the entropy production associated
with the ``average'' input state $\rho^{Q} = \sum_{k} p_{k} \rho^{Q}_{k}$,
which equals the entropy of the average final environment state
$\rho^{E'}$.

The eavesdropper Eve will try to infer the preparation $\rho^{Q}_{k}$ by
examining the state of her measuring apparatus---that is, by trying
to distinguish the various environment states $\rho^{E'}_{k}$.  Denote
Alice's preparation, and thus the final environment state produced by
that preparation, by the random variable $X$ and the reading on Eve's
measuring apparatus by $Y$.  Then a theorem of Kholevo 
\cite{kholevo} limits the
mutual information $I(X:Y)$, which is the amount of information about
$X$ that Eve obtains from a knowledge of $Y$.  This limit is
\begin{eqnarray}
	I(X:Y) & \leq & S(\rho^{E'}) - \sum_{k} p_{k} S(\rho^{E'}_{k}) 
			\nonumber \\
		& = & S_{e} - \sum_{k} p_{k} S_{e,k}    \\
		& \leq & S_{e} .
\end{eqnarray}
(If the eavesdropper Eve only has access to part of the environment system
$E$, then she will be able to do no better, and $I(X:Y)$ will still be
bounded in this way.)

Thus, the entropy production associated with the ensemble of input states and
the evolution superoperator $\superop^{Q}$, both of which can be
determined in principle from repeated use of the channel $Q$, 
limits the amount of information that any eavesdropper might obtain 
about the input.  Put another way, any process by which the eavesdropper
obtains information about the channel system $Q$ disturbs the system,
leaving traces in the evolution superoperator $\superop^{Q}$.  The
disturbance produced by the eavesdropper (and other interactions with
the environment) is characterized by the entropy production $S_{e}$.

\section{The quantum Fano inequality}

\subsection{Classical theorem}

In classical information theory, there is a simple relation between
the noise in a channel and probability of error in that channel \cite{cover}.  
This relation is Fano's inequality.  We will derive an analogous
quantum relation.

Let $X$ be a classical random variable representing the input of a 
noisy channel, and suppose that $X$ can take on up to $N$ different
values.  The output of the noisy channel is represented by the random
variable $Y$.  The channel itself is represented by the conditional
probabilities $p(y_{k}|x_{j})$ of an output value 
$y_{k}$ given an input value $x_{j}$.
These probabilities, together with the input probability distribution
$p(x_{j})$, characterize the situation.
The receiver makes an estimate $\hat{X}$ of the input
$X$ based only on the channel output $Y$.  The probability of error
$P_{E}$ is the total likelihood that $\hat{X} \neq X$.

Fano's inequality (in its stronger form) states that
\begin{equation}
   h(P_{E}) + P_{E} \log (N-1) \geq H(X|Y)  \label{stronfano}
\end{equation}
where $h(P_{E}) = - P_{E} \log P_{E} - (1-P_{E}) \log P_{E}$ and
$H(X|Y)$ is the Shannon conditional entropy of $X$ given $Y$.  
$H(X|Y)$, the average residual information uncertainty
about the input given the output, is a measure of the noise in
the channel.  $H(X|Y) = 0$ for a noiseless channel, in which the input
$X$ can be exactly determined by the output $Y$.
Noting that $h(P_{E}) \leq 1$ (since our logarithms are base 2),
we can derive a simpler but slightly weaker form of Fano's inequality.
\begin{equation}
	1 + P_{E} \log N  >  H(X|Y) .   \label{weakfano}
\end{equation}
Fano's inequality is used to prove the ``weak converse'' of the classical
noisy coding theorem, which states that information cannot be sent
at a rate greater than the channel capacity with arbitrarily low probability
of error \cite{cover}.

\subsection{Quantum theorem}

We now turn to the quantum problem.  As before, we suppose that the system
$RQ$ is initially in the entangled state $\ket{\Psi^{RQ}}$, and that 
$Q$ is subjected to an evolution described by $\superop^{Q}$.  The reference
system $R$ is isolated and has trivial dynamics described by $\unity^{R}$.
The dimensions of $\hilbert{Q}$ and $\hilbert{R}$ are both finite and
equal to $d$.  After the evolution, the system is described by a joint 
state $\rho^{RQ'}$.

Now suppose that we subject the final state $\rho^{RQ'}$ to a 
measurement of a complete
ordinary observable on the system $RQ$, which is described by a basis
of $d^{2}$ orthogonal states for $RQ$.  Let the random variable $X$
represent the outcome of this measurement.  Then we know (from 
Equation~\ref{entineq}) that
\begin{displaymath}
	S_{e} = S(\rho^{RQ'}) \leq H(X) .
\end{displaymath}
Further suppose that one of these
basis vectors is chosen to be the original state $\ket{\Psi^{RQ}}$.
Then the fidelity $F_{e} = \bra{\Psi^{RQ}} \rho^{RQ'} \ket{\Psi^{RQ}}$
is just the probability of this outcome.  Given this probability, the
largest possible value of $H(X)$ would occur when all of the $d^{2} - 1$
other outcomes have equal probability.  Then
\begin{eqnarray*}
	\max H(X)  & = &  - F_{e} \log F_{e} 
				- (d^{2}-1) \frac{1-F_{e}}{d^{2}-1}
					\log \frac{1-F_{e}}{d^{2}-1}  \\
		   & = &  - F_{e} \log F_{e} - (1-F_{e}) \log (1-F_{e})
					+ (1-F_{e}) \log (d^{2} - 1) .
\end{eqnarray*}
Therefore we can conclude that
\begin{equation}
	h(F_{e}) + (1-F_{e}) \log (d^{2} - 1)  \geq  S_{e} . 
		\label{strongqfano}
\end{equation}
This is our quantum version of the Fano inequality, relating the entanglement
fidelity $F_{e}$ with the entropy production $S_{e}$.  Although we have 
made use of the reference system $R$ in deriving this inequality, both
$F_{e}$ and $S_{e}$ have meanings that are intrinsic to the system $Q$.

As before, we can give a slightly weaker form of the inequality:
\begin{equation}
	1 + 2 (1-F_{e}) \log d  \geq  S_{e} .
\end{equation}
It is instructive to compare the form of this equation to that of equation
\ref{weakfano}.  The number $N$ of possible input states is analogous
the dimension $d$ of $\hilbert{Q}$.
The probability of error $P_{E}$ roughly corresponds 
$1 - F_{e}$, the amount by which the final entangled state fails to 
correspond to the initial one.  The noise term $H(X|Y)$ is replaced by
the entropy production $S_{e}$.  Finally, a factor of 2 appears in the
error term in the quantum case---which in fact corresponds to replacing
$N$ by $d^{2}$, the dimension of $\hilbert{Q} \otimes \hilbert{R}$.

We can strengthen the quantum Fano inequality in a number of ways.
First, if the reference system $R$ has a Hilbert space of dimension
$d_{R} < d$, the quantity $d^{2}$ can be replaced by the product $d_{R} d$.
The required dimension $d_{R}$ is in fact just the dimension of the 
subspace that supports $\rho^{R}$, and so $d_{R} \leq d$ even if $R$
is much larger than $Q$.  Since we wish to consider $F_{e}$ and $S_{e}$
to be quantities intrinsic to $Q$, though, we will simply adopt $d_{R} = d$.

Finally, we note that the fidelity $F_{e}$ can
be lowered by {\em internal dynamics} of $Q$ as well as by information
exchange with the environment.  To take this into account, we could allow
the final state of the system to be ``processed'' via any unitary 
transformation $U^{Q}$ on $Q$, and define
\begin{equation}
	\hat{F}_{e} = \max_{U^{Q}}  \bra{\Psi^{RQ}} 
			\left ( 1^{R} \otimes U^{Q} \right ) \rho^{RQ'} 
			\left ( 1^{R} \otimes U^{Q} \right )^{\dagger}
			\ket{\Psi^{RQ}} .
\end{equation}
($\hat{F}_{e}$ is also independent of the particular purification 
for $\rho^{Q}$, and is thus an quantity intrinsic to $Q$.)
Clearly $\hat{F}_{e} \geq F_{e}$.  A derivation very similar to the
one we have given allows us to replace $F_{e}$ by $\hat{F}_{e}$ in
equation \ref{strongqfano}, obtaining
\begin{eqnarray}
	h(\hat{F}_{e}) + (1-\hat{F}_{e}) \log (d^{2} - 1)  & \geq & S_{e} \\
	1 + 2 (1 - \hat{F}_{e}) \log d & \geq & S_{e} .
\end{eqnarray}

We could further extend this by allowing $Q$ to be subjected to a second
arbitrary completely positive map after $\superop^{Q}$, and obtain a similar
relation.  However, in this case the relevant entropy production 
$\hat{S}_{e}$ would be that due to the total evolution, both $\superop^{Q}$
and the subsequent ``processing''.  Since it is
possible that $\hat{S}_{e} < S_{e}$, we do not obtain a useful general
relation.  (This is precisely what happens
in quantum error-correcting codes, as explained below.)

\section{Remarks}

One possible application of entanglement fidelity and entropy production
is in the study of non-ideal quantum computers \cite{qcompute}.
In a typical state of a quantum computer, the different parts of the computer
are in a highly entangled state.  The elements of the computer's memory 
must maintain their states in such a fashion that this entanglement is
preserved.  The considerations in these notes are thus particularly suited
to studying the effects of noise and decoherence in this context.

What we have found is that the capability of a system $Q$ to preserve its
entanglement with some other system $R$ can be determined from the initial
state and the dynamics of $Q$ itself.  Destruction or distortion of 
entanglement, and information exchange with the environment, leave 
distinct traces in the dynamics of the system itself.  We can characterize
these by the entanglement fidelity $F_{e}$ and the entropy production $S_{e}$.

$F_{e}$ is properly thought of, not as the fidelity of one state with 
another (though it can be given that interpretation by including a 
reference system $R$), but as the fidelity of a {\em process} given by
the input state $\rho^{Q}$ and the system dynamics $\superop^{Q}$.  
$F_{e}$ does not just measure how {\em well} the state of $Q$ is 
preserved by $\superop^{Q}$, but also how {\em coherently}.  If 
the input state is a pure state, these amount to the same thing; but
otherwise, $F_{e}$ is a stronger measure of the amount of disturbance
the state experiences.

$S_{e}$ is also properly thought of, not as the entropy of some state, but 
as the entropy associated with the dynamical
process given by $\rho^{Q}$ and $\superop^{Q}$.  Information exchange with
the environment, even if it does not change the entropy of either the
system $Q$ or the environment $E$, can lead to non-zero entropy production
$S_{e}$.  Entropy production is therefore a clearer measure of this
exchange than the changes in entropy of either system.

The relationship between $F_{e}$ and $S_{e}$ amounts to a quantum Fano
inequality, connecting the information exchange with the environment
to the disturbance of the state.  This illustrates very clearly a
general principle:  {\em In quantum information theory, noise is exactly
information exchange with an external system.}  In a classical system,
information can be ``leaked'' into the environment with arbitrarily
little disturbance to he system---the environment can simply make a 
copy of the information, leaving the original intact within the system.  
But quantum information cannot be copied.  Any departure of information
into the environment necessarily yields an irreducible disturbance of 
the system.  (This is the fundamental idea behind quantum cryptography.)
The departing information leaves its ``footprints'' behind in the
entropy production $S_{e}$ and associated imperfect entanglement 
fidelity $F_{e}$. 

These ideas shed an interesting light on the recently discovered 
quantum error-correcting codes \cite{qerror}.  In these codes, input
quantum states are represented by massively entangled states of 
a system $Q$ composed of many qubits:  $Q = Q_{1} \cdots Q_{n}$.  
The environment is assumed to act independently on these 
systems, which in our language corresponds to the requirement that
the evolution superoperator for the system $Q$ factorizes:
\begin{displaymath}
	\superop^{Q} =
		\superop^{Q_{1}} \otimes \cdots \otimes \superop^{Q_{2}}  .
\end{displaymath}
The resulting state is then subjected to a second process, which 
typically involves an incomplete measurement on $Q$ 
followed by a unitary evolution (which depends
on the measurement result).  Under certain circumstances, the original
state of the system may be restored with very high fidelity.

The action of the channel and the subsequent restoration process 
of the sequence of qubits can be written as a single superoperator for
$Q_{1} \cdots Q_{n}$.  Since the fidelity of this combined process is
high, we can conclude, rather surprisingly, that the total entropy 
production is quite low.  At first this seems paradoxical, since the
individual entropy productions of the ``noise'' process and the 
restoration measurement may both be high.  

But this is not too difficult to understand.  Let $E$ represent the 
environment system that interacts with the qubits during the ``noise''
stage, and let $M$ represent the apparatus that performs the restoration
process.  To begin with, we might imagine that $E$ and $M$ are in pure
states.  After $Q$ interacts with $E$ (and thus exchanges information),
the state of $QE$ becomes entangled.  In the second stage, $M$ interacts
and exchanges information with $Q$, and the entanglement of $Q$ with the
rest of the world is reduced---it is passed to $M$.  At the end of the
process, both $Q$ and the ``rest of the world'' $EM$ are in near-pure states,
but $E$ and $M$ have now become entangled.

Thus, the process of quantum error-correction can be thought of as a process
of passing entanglement (produced by a previous interaction with the
environment) to the apparatus, in such a way that the entropy production
for the total process (noise followed by restoration) on $Q$ is very low.
If $S_{e}$ is very low, then the overall dynamics for $Q$ is nearly unitary, 
so that the original state of $Q$ can be approximately recovered.
It is not yet known under what general circumstances, and to what fidelity,
this can be accomplished.

\subsection*{Acknowledgments}

The author is indebted to many people for extensive conversations about
the issues discussed in this paper, including H. Barnum, C. H. Bennett, 
C. M. Caves, I. Chuang, A. Ekert, C. A. Fuchs, E. H. Knill, R. Jozsa, 
R. Laflamme, J. Smolin, M. D. Westmoreland, W. K. Wootters, and W. H. Zurek.
He also wishes to acknowledge the hospitality and support of the
Theoretical Astrophysics group (T-6) at Los Alamos National
Laboratory during 1995--96.

\renewcommand{\thesection}{\Alph{section}}
\setcounter{section}{1}

\section*{Appendix:  Representation theorems}

\subsection{Index states and relative states}

In this appendix we will use some of the ideas from the main paper to 
show that any trace-preserving, completely positve linear map has
both an operator-sum representation and a unitary representation.
This derivation is somewhat more direct than that found in \cite{repthms}.
We will also suggest a useful characterization of all such representations.

Suppose $R$ and $Q$ are quantum systems with $\dim \hilbert{R} =
\dim \hilbert{Q} = d$, and let $\ket{\alpha^{R}_{k}}$ and 
$\ket{\beta^{Q}_{k}}$ be orthonormal basis vectors for $\hilbert{R}$
and $\hilbert{Q}$.  We can write down a ``maximally entangled'' pure
state of $RQ$:
\begin{displaymath}
    \ket{\Psi^{RQ}} = \frac{1}{\sqrt{d}} \sum_{k} 
        \ket{\alpha^{R}_{k}} \otimes \ket{\beta^{Q}_{k}} .
\end{displaymath}
It will be convenient to consider instead the non-normalized vector
\begin{displaymath}
    \ket{\tilde{\Psi}^{RQ}} = \sqrt{d} \ket{\Psi^{RQ}} = 
      \sum_{k} \ket{\alpha^{R}_{k}} \otimes \ket{\beta^{Q}_{k}} .
\end{displaymath}
(Using $\ket{\tilde{\Psi}^{RQ}}$ rather than $\ket{\Psi^{RQ}}$ will
eliminate some factors of $\sqrt{d}$ in our expressions.)

For every state $\ket{\zeta^{R}}$ of $R$ there is a unique state
$\ket{\xi^{Q}}$ such that
\begin{eqnarray*}
    \frac{1}{\sqrt{d}} \ket{\xi^{Q}} & = &
        \amp{\zeta^{R}}{\Psi^{RQ}}      \\
    \ket{\xi^{Q}}   & = &   \amp{\zeta^{R}}{\tilde{\Psi}^{RQ}} .    
\end{eqnarray*}
The relation between $\ket{\zeta^{R}}$ and $\ket{\xi^{Q}}$ is a one-to-one
correspondence.  We call $\ket{\xi^{Q}}$ the {\em relative state} in $Q$
to $\ket{\zeta^{R}}$, and we call $\ket{\zeta^{R}}$ the {\em index state}
in $R$ that yields $\ket{\xi^{Q}}$.

Given a state $\ket{\phi^{Q}}$, let us denote the associated index state
in $R$ by $\ket{\phi^{\ast R}}$.  We can give a simple prescription for 
finding $\ket{\phi^{\ast R}}$ from $\ket{\phi^{Q}}$.  Suppose
\begin{displaymath}
    \ket{\phi^{Q}} = \sum_{k} c_{k} \ket{\beta^{Q}_{k}} .
\end{displaymath}
Then 
\begin{displaymath}
    \ket{\phi^{\ast R}} = \sum_{k} c_{k}^{\ast} \ket{\alpha^{R}_{k}},
\end{displaymath}
as can be easily seen:
\begin{eqnarray*}
    \amp{\phi^{\ast R}}{\tilde{\Psi}^{RQ}}
      & = & \sum_{k l} c_{k} 
        \amp{\alpha^{R}_{k}}{\alpha^{R}_{l}} \ket{\beta^{Q}_{l}} \\
      & = & \sum_{k} c_{k} \ket{\beta^{Q}_{k}} \\
      & = & \ket{\phi^{Q}} .
\end{eqnarray*}
It is also clear that
\begin{eqnarray*}
    \lefteqn{\proj{\phi^{\ast R}} \otimes \proj{\phi^{Q}} =} \\
    &  &    \left ( \proj{\phi^{\ast R}} \otimes 1^{Q} \right )
            \proj{\tilde{\Psi}^{RQ}} 
        \left ( \proj{\phi^{\ast R}} \otimes 1^{Q} \right ),
\end{eqnarray*}
a relation that will be useful later on.

The function that takes $\ket{\phi^{Q}}$ to $\ket{\phi^{\ast R}}$ is
conjugate linear.  If $\ket{\phi^{Q}} = a_{1} \ket{\phi^{Q}_{1}}
+ a_{2} \ket{\phi^{Q}_{2}}$, then
\begin{eqnarray*}
   \ket{\phi^{\ast R}}
    & = &   a_{1}^{\ast} \ket{\phi^{\ast R}_{1}} 
        + a_{2}^{\ast} \ket{\phi^{\ast R}_{2}}  \\
   \bra{\phi^{\ast R}}
    & = &   a_{1} \bra{\phi^{\ast R}_{1}} 
        + a_{2} \bra{\phi^{\ast R}_{2}} .
\end{eqnarray*}

\subsection{Operator-sum representations}

Let $\superop^{Q}$ be the trace-preserving, completely positive linear
map that describes the dynamical evolution of the system $Q$.  Since 
$\superop^{Q}$ is completely positive, any trivial extension of it is
positive; in particular, the superoperator $\unity^{R} \otimes \superop^{Q}$
is positive.  Thus, the state
\begin{displaymath}
    \rho^{RQ'} = \unity^{R} \otimes \superop^{Q} 
            \left ( \proj{\Psi^{RQ}} \right )
\end{displaymath}
is a positive operator, as is
\begin{displaymath}
    D^{RQ'} = d \, \rho^{RQ'}  =  \unity^{R} \otimes \superop^{Q} 
            \left ( \proj{\tilde{\Psi}^{RQ}} \right ) .
\end{displaymath}
Of course, $\rho^{RQ'}$ has unit trace, so it is a normalized density
operator, while $\tr \, D^{RQ'} = d$.

The operation of realizing a state of $Q$ via choosing an index state of
$R$ commutes with the dynamical operation given by $\unity^{R} \otimes
\superop^{Q}$.  In other words, if we wish to write down the final state
$\rho^{Q'} = \superop^{Q} (\rho^{Q})$, where $\rho^{Q} = \proj{\phi^{Q}}$,
we can either apply the index state $\ket{\phi^{\ast R}}$ to 
$\ket{\tilde{\Psi}^{RQ}}$ and then apply $\superop^{Q}$, or else we
can apply the extended superoperator $\unity^{R} \otimes \superop^{Q}$
to the joint state and then apply the index state, thus:
\begin{displaymath}
    \rho^{Q'} = \bra{\phi^{\ast R}} D^{RQ'} \ket{\phi^{\ast R}} .
\end{displaymath}
This makes sense on physical grounds.  A measurement of an observable on $R$
involves a completely different system than the dynamical evolution of $Q$,
and the two operations might take place arbitrarily far apart.  
The time order of the two should irrelevant to the result.

A more formal argument runs as follows.  Let $\Phi^{R}$ be the superoperator
(i.e., a linear map on operators on $\hilbert{R}$) associated with 
multiplication by $\proj{\phi^{\ast R}}$ on both sides.  That is, if $T^{R}$
is an operator on $\hilbert{R}$, then $\Phi^{R} ( T^{R} )  =
\proj{\phi^{\ast R}} T^{R} \proj{\phi^{\ast R}}$.  The superoperator 
$\Phi^{R} \otimes \unity^{Q}$ (which is just multiplication on both sides 
by $\proj{\phi^{\ast R}} \otimes 1^{Q}$) obviously commutes with the
dynamical superoperator $\unity^{R} \otimes \superop^{Q}$.  Therefore,
\begin{eqnarray*}
\lefteqn{\Phi^{R} \otimes \unity^{Q} \left ( D^{RQ'} \right ) }  \\
  & = & \Phi^{R} \otimes \unity^{Q} \left ( \unity^{R} \otimes \superop^{Q}
        \left ( \proj{\tilde{\Psi}^{RQ}} \right ) \right ) \\
  & = & \unity^{R} \otimes \superop^{Q} \left ( \Phi^{R} \otimes \unity^{Q}
        \left ( \proj{\tilde{\Psi}^{RQ}} \right ) \right ) \\
  & = & \unity^{R} \otimes \superop^{Q} \left (
        \left( \proj{\phi^{\ast R}} \otimes 1^{Q} \right )
        \proj{\tilde{\Psi}^{RQ}} 
        \left( \proj{\phi^{\ast R}} \otimes 1^{Q} \right ) \right ) \\
  & = & \unity^{R} \otimes \superop^{Q} \left ( 
        \proj{\phi^{\ast R}} \otimes \proj{\phi^{Q}} \right ) \\
  & = & \proj{\phi^{\ast R}} \otimes \rho^{Q'} .
\end{eqnarray*}
From this we can see that
\begin{displaymath}
    \rho^{Q'} = \superop^{Q} \left ( \proj{\phi^{Q}} \right ) =
        \bra{\phi^{\ast R}} D^{RQ'} \ket{\phi^{\ast R}} 
\end{displaymath}
as we wished to show.

The operator $D^{RQ'}$ is positive; thus, we can find 
a set of vectors $\ket{\tilde{\mu}^{RQ'}}$ such that
\begin{displaymath}
    D^{RQ'} = \sum_{\mu} \proj{\tilde{\mu}^{RQ'}} .
\end{displaymath}
These vectors, for example, might be constructed from the eigenvectors
of $D^{RQ'}$, normalized by their eigenvalues; but there are many such
decompositions.  In fact, it is easy to see that the 
$\ket{\tilde{\mu}^{RQ'}}$ vectors are simply related to the
representation of $\rho^{RQ'}$ by an ensemble of pure states.  
That is, given such a representation
\begin{displaymath}
    \rho^{RQ'} = \sum_{\mu} p_{\mu} \proj{\psi^{RQ'}_{\mu}}
\end{displaymath}
we can simply set $\ket{\tilde{\mu}^{RQ'}} = \sqrt{p_{\mu} d} 
\ket{\psi^{RQ'}_{\mu}}$.

It is also clear that there is a decomposition of $D^{RQ'}$ with
no more than $d^{2}$ vectors $\ket{\tilde{\mu}^{RQ'}}$, since the
dimension of the space $\hilbert{R} \otimes \hilbert{Q}$ is $d^{2}$.

Here comes the essential trick.  Define the operator $A^{Q}_{\mu}$ by
\begin{displaymath}
    A^{Q}_{\mu} \ket{\phi^{Q}} 
        = \amp{\phi^{\ast R}}{\tilde{\mu}^{RQ'}}
\end{displaymath}
for each state $\ket{\phi^{Q}}$ of $Q$.  Because of the conjugate linear
relation between $\ket{\phi^{Q}}$ and $\ket{\phi^{\ast R}}$, each 
$A^{Q}_{\mu}$ thus defined is a perfectly good linear operator
on $\hilbert{Q}$.  Furthermore,
\begin{eqnarray*}
\sum_{\mu} A^{Q}_{\mu} \proj{\phi^{Q}} {A^{Q}_{\mu}}^{\dagger}
    & = &   \sum_{\mu} \amp{\phi^{\ast R}}{\tilde{\mu}^{RQ'}}
        \amp{\tilde{\mu}^{RQ'}}{\phi^{\ast R}} \\
    & = &   \bra{\phi^{\ast R}} D^{RQ'} \ket{\phi^{\ast R}} \\
    & = &   \superop^{Q} \left ( \proj{\phi^{Q}} \right ) .
\end{eqnarray*}
We have thus derived an operator-sum representation for the completely
positive map $\superop^{Q}$ for all pure input states $\proj{\phi^{Q}}$.
Extending this to mixed state inputs is trivial, of course, since every
mixed state is a linear (convex) combination of pure states.  We can
further see that each completely positive map $\superop^{Q}$ has
an operator-sum representation with no more than $d^{2}$ terms.

We also find that, for our operator-sum representation for $\superop^{Q}$,
\begin{eqnarray*}
\sum_{\mu} \bra{\phi^{Q}} A^{Q}_{\mu} {A^{Q}_{\mu}}^{\dagger} 
        \ket{\phi^{Q}} 
    & = & \tr \, \sum_{\mu} A^{Q}_{\mu} 
        \proj{\phi^{Q}} {A^{Q}_{\mu}}^{\dagger} \\
    & = & \tr \, \rho^{Q'} \\
    & = & 1
\end{eqnarray*}
since $\superop^{Q}$ is trace-preserving by assumption.  Since this is
true for all states $\ket{\phi^{Q}}$, including the eigenstates of the
positive operator $\sum_{\mu} {A^{Q}_{\mu}}^{\dagger} A^{Q}_{\mu}$,
we conclude that
\begin{displaymath}
    \sum_{\mu} {A^{Q}_{\mu}}^{\dagger} A^{Q}_{\mu} = 1^{Q} .
\end{displaymath}

\subsection{Unitary representations}

Having derived an operator-sum representation for $\superop^{Q}$, it is
easy to arrive at a unitary representation.  Add an extra
quantum system $E$ and write down a purification 
$\ket{\tilde{\Upsilon}^{RQE'}}$ for $D^{RQ'}$ as follows:
\begin{displaymath}
    \ket{\tilde{\Upsilon}^{RQE'}} = \sum_{\mu} \ket{\tilde{\mu}^{RQ'}}
            \otimes \ket{\epsilon^{E}_{\mu}}
\end{displaymath}
for an orthonormal set of vectors $\ket{\epsilon^{E}_{\mu}}$ in 
$\hilbert{E}$.  (Again, finding a purification for $D^{RQ'}$ is
equivalent to finding a purification for $\rho^{RQ'}$, but it is
slightly easier to work with the non-normalized states.)
We note that we require no more than $d^{2}$ dimensions in $\hilbert{E}$
to construct this purification, since there are decompositions of
$D^{RQ'}$ with no more than $d^{2}$ vectors $\ket{\tilde{\mu}^{RQ'}}$.
Fix some state $\ket{0^{E}}$ of $E$.  
We can define an operator $U^{QE}$ on a subspace of
$\hilbert{Q} \otimes \hilbert{E}$ by
\begin{eqnarray*}
    U^{QE} \left ( \ket{\phi^{Q}} \otimes \ket{0^{E}} \right )
        & = & \amp{\phi^{\ast R}}{\tilde{\Upsilon}^{RQE'}} \\
        & = & \sum_{\mu} \amp{\phi^{\ast R}}{\tilde{\mu}^{RQ'}}
            \otimes \ket{\epsilon^{E}_{\mu}}  \\
        & = & \sum_{\mu} A^{Q}_{\mu} \ket{\phi^{Q}} \otimes
            \ket{\epsilon^{E}_{\mu}} \\
        & = & \ket{\Phi^{QE'}} 
\end{eqnarray*}
for all $\ket{\phi^{Q}}$ in $\hilbert{Q}$.  Once again, the conjugate
linear relation of index state and relative state guarantees that this
is a linear operator.  Furthermore, given two states $\ket{\phi^{Q}_{1}}$
and $\ket{\phi^{Q}_{2}}$,
\begin{eqnarray*}
  \amp{\Phi^{QE'}_{1}}{\Phi^{QE'}_{2}}
    & = &   \amp{\tilde{\Upsilon}^{RQE'}}{\phi^{\ast R}_{1}}
        \amp{\phi^{\ast R}_{2}}{\tilde{\Upsilon}^{RQE'}}  \\
    & = &   \sum_{\mu \nu} \bra{\phi^{Q}_{1}} {A^{Q}_{\mu}}^{\dagger} 
            A^{Q}_{\nu} \ket{\phi^{Q}_{2}}
            \amp{\epsilon^{Q}_{\mu}}{\epsilon^{E}_{\nu}} \\
    & = &   \sum_{\mu} \bra{\phi^{Q}_{1}} {A^{Q}_{\mu}}^{\dagger} 
            A^{Q}_{\mu} \ket{\phi^{Q}_{2}} \\
    & = &   \amp{\phi^{Q}_{1}}{\phi^{Q}_{2}} .
\end{eqnarray*}
The operator $U^{QE}$ preserves inner products on this subspace of states;
it can therefore be extended to a unitary operator on the entire space
$\hilbert{Q} \otimes \hilbert{E}$.

Thus, we have a unitary representation for $\superop^{Q}$:
\begin{eqnarray*}
\lefteqn{\tr_{E} U^{QE} \left ( \proj{\phi^{Q}} \otimes \proj{0^{E}}
        \right ) {U^{QE}}^{\dagger} }  \\
    & = & \tr_{E} \sum_{\mu \nu}  \left ( A^{Q}_{\mu} \proj{\phi^{Q}}
        {A^{Q}_{\nu}}^{\dagger} \right ) \otimes 
        \ket{\epsilon^{E}_{\mu}} \bra{\epsilon^{E}_{\nu}} \\
    & = & \sum_{\mu \nu}  \left ( A^{Q}_{\mu} \proj{\phi^{Q}}
        {A^{Q}_{\nu}}^{\dagger} \right )
        \amp{\epsilon^{E}_{\nu}}{\epsilon^{E}_{\mu}} \\
    & = & \sum_{\mu} A^{Q}_{\mu} \proj{\phi^{Q}}
        {A^{Q}_{\nu}}^{\dagger} \\
    & = & \superop^{Q} \left ( \proj{\phi^{Q}} \right ) .
\end{eqnarray*}
Once again, we can extend this unitary representation to mixed state
inputs, since these are linear (convex) combinations of pure states.

\subsection{Remarks}

In the above arguments, we arrived at an operator-sum representation
for $\superop^{Q}$ by a decomposition of $D^{RQ'}$, that is, by a 
pure state ensemble for $\rho^{RQ'}$.  It is also easy to see that
every operator-sum representation for $\superop^{Q}$, when extended
and applied to $\ket{\Psi^{RQ}}$, will yield such a decomposition.
(Simply define $\ket{\tilde{\mu}^{RQ'}} = (1^{R} \otimes A^{Q}_{\mu})
\ket{\Psi^{RQ}}$.)  Thus, the operator-sum representations for 
$\superop^{Q}$ are in a one-to-one correspondence with the pure 
state ensembles for $\rho^{RQ'}$.

Similarly, we obtained a unitary representation for $\superop^{Q}$ by
finding a purification for $D^{RQ'}$, or equivalently, for $\rho^{RQ'}$.
But every unitary representation will be associated with such a 
purification, because the initial total state 
$\ket{\Psi^{RQ}} \otimes \ket{0^{E}}$ of $RQE$ will evolve unitarily
to a pure state, from which the state $\rho^{RQ'}$ is obtained by
a partial trace over $E$.  Now, any such purification of $\rho^{RQ'}$
can be obtained from any other by means of
a unitary transformation that acts on $\hilbert{E}$,
which corresponds to an internal rotation of the environment system $E$
that acts {\em after} the interaction of $Q$ and $E$.

The non-uniqueness of the operator-sum representation and the unitary
representations are related, since every pure state ensemble for
$\rho^{RQ'}$ can be realized by fixing a purification $\ket{\Upsilon^{RQE'}}$
and choosing a complete ordinary measurement for $E$ (i.e., an orthonormal 
basis for $\hilbert{E}$).  Equivalently, 
we might fix a measurement basis for 
$\hilbert{E}$ and a particular purification.
A change of representation in each case will be associated 
with a unitary matrix corresponding to a rotation in $\hilbert{E}$.
That is, suppose that for all $\rho^{Q}$,
\begin{displaymath}
	\superop^{Q}(\rho^{Q}) =
		\sum_{\mu} A^{Q}_{\mu} \rho^{Q} {A^{Q}_{\mu}}^{\dagger}
	    =   \sum_{\nu} B^{Q}_{\nu} \rho^{Q} {B^{Q}_{\nu}}^{\dagger}
\end{displaymath}
so that the $A^{Q}_{\mu}$ and the $B^{Q}_{\nu}$ operators both form
operator-sum representations for $\superop^{Q}$.  Then there is a 
unitary matrix $U_{\mu \nu}$ so that
\begin{displaymath}
	A^{Q}_{\mu}  =  \sum_{\nu} U_{\mu \nu} B^{Q}_{\nu} .
\end{displaymath}
(Note that we may have to extend one operator-sum representation by a
finite number of zero operators so that the two representations have
the same number of operators.)
The matrix $U_{\mu \nu}$ is in fact the matrix that relates two
different bases in $E$, corresponding to two purifications related,
in the sense outlined above, to the two operator-sum representations.

\pagebreak


\begin{thebibliography}{99}
%
\bibitem{chb-phystoday}  C. H. Bennett, {\em Physics Today} {\bf 48},
	24 (1995).
%
\bibitem{shannon}  C. E. Shannon, {\em Bell System Tecyhnical Journal}
	{\bf 27}, 379 (1948).
%
\bibitem{qcoding}  R. Jozsa and B. Schumacher, {\em Journal of Modern Optics}
	{\bf 41}, 2343 (1994).  B. Schumacher {\em Physical Review A}
	{\bf 51}, 2738 (1995).  H. Barnum, C. A. Fuchs, R. Jozsa, and
	B. Schumacher, ``General fidelity limit for quantum channels'',
	preprint.
%
\bibitem{qfidelity}  R. Jozsa, {Journal of Modern Optics} {\bf 41}, 2315
	(1995).
%
\bibitem{entang}  C. H. Bennett and S. J. Wiesner, {\em Physical Review
	Letters} {\bf 68}, 3121 (1992).  C. H. Bennett, G. Brassard,
	C. Cr\'{e}peau, R. Jozsa, A. Peres, and W. K. Wootters, 
	{\em Physical Review Letters} {\bf 69}, 2881 (1992).
%
\bibitem{qerror}  P. W. Shor, {\em Physical Review A} {\bf 52}, 2493 (1995).
	A. R. Calderbank and P. W. Shor, ``Good quantum error-correcting
	codes exist'', preprint.  A. Steane, ``Multiple particle interference
	and quantum error correction'', preprint.  R. Laflamme, C. Miquel,
	J. P. Paz, and W. H. Zurek, ``Perfect quantum error correction code'',
	preprint.
%
\bibitem{purify}  C. H. Bennett, G. Brassard, S. Popescu, B. Schumacher,
	J. A. Smolin, and W. K. Wootters, {\em Physical Review Letters}
	{\bf 76}, 722 (1996).
%
\bibitem{cpmaps}  W. F. Stinespring, {\em Proceedings of the American
	Mathematical Society} {\bf 6}, 211 (1955).  K. Kraus, {\em
	Annals of Physics} {\bf 64}, 311 (1971).  
%
\bibitem{repthms}  K. Hellwig and K. Kraus, {\em Communications in 
	Mathematical Physics} {\bf 16}, 142 (1970).  M.-D. Choi,
	{\em Linear Algebra and Its Applications} {\bf 10}, 285 (1975).
	K. Kraus, {\em States, Effects, and Operations:  Fundamental 
	Notions of Quantum Theory} (Springer-Verlag, Berlin, 1983).
%
\bibitem{ensembles}  L. P. Hughston, R. Jozsa, and W. K. Wooters, 
	{\em Physics Letters A} {\bf 183}, 14 (1993).
%
\bibitem{relstate}  H. Everett III, {\em Reviews of Modern Physics}
	{\bf 29}, 454 (1957).
%
\bibitem{posops}  C. W. Helstrom, {\em Quantum Detection and Estimation
	Theory} (Academic Press, New York, 1976).  A. Peres, {\em Foundations
	of Physics} {\bf 20}, 1441 (1990).
%
\bibitem{vonneumann}  J. von Neumann, {\em Mathematical Foundations of
	Quantum Mechanics}, translated by E. T. Beyer (Princeton University
	Press, Princeton, 1955).
%
\bibitem{wehrl}  A. Wehrl, {\em Reviews of Modern Physics} {\bf 50}
	221 (1978).
%
\bibitem{cover}  T. M. Cover and J. A. Thomas, {\em Elements of Information
	Theory} (Wiley, New York, 1991).
%
\bibitem{qcrypto}  C. H. Bennett and G. Brassard, {\em Proceedings of the
	IEEE Conference on Computers, Systems, and Signal Processing,
	Bangalore, India} (IEEE, New York, 1984), p. 175.  C. H. Bennett,
	F. Bessette, G. Brassard, L. Salvail, and J. Smolin, 
	{\em Journal of Cryptology} {\bf 5}, 3 (1992).
%
\bibitem{qcrypto-entang}  A. K. Ekert, {\em Physical Review Letters}
	{\bf 67}, 661 (1991).
%
\bibitem{kholevo}  A. S. Kholevo, {\em Problemy Peredachi Informatsii} 
	{\bf 9}, 3 (1973); translated in {\em Problems of Information 
	Transmission (USSR)} {\bf 9}, 177 (1973).  C. Caves and C. Fuchs,
	{\em Physical Review Letters} {\bf 73}, 3047 (1994).  B. Schumacher,
	M. D. Westmoreland, and W. K. Wootters, ``Limitation on the amount
	of accessible information in a quantum channel'', {\em Physical
	Review Letters}, to appear (1996).
%
\bibitem{qcompute}  S. Lloyd, {\em Scientific American} {\bf 273}, 140
	(1995).  A. Ekert and R. Jozsa, ``Notes on Shor's efficient
	algorithm for factoring on a quantum computer'', {\em Reviews of
	Modern Physics}, to appear (1996).
%
\end{thebibliography}
\end{document}